\documentclass[lettersize,journal]{IEEEtran}

\usepackage[noadjust]{cite}
\usepackage[english]{babel}
\usepackage{blindtext}
\usepackage{booktabs} 
\usepackage{epsfig,endnotes, enumerate}
\usepackage{subfigure}
\usepackage{graphicx}
\usepackage{url}
\usepackage{color}
\usepackage{balance}
\usepackage{amssymb,amsmath, listings, verbatim, float}
\usepackage[linesnumbered, ruled, vlined]{algorithm2e}
\usepackage{mathtools, comment}
\usepackage{multirow}
\usepackage{etoolbox}

\newcommand{\name}{LoRaIN}

\definecolor{blue}{rgb}{0.0,0.0,1}
\newcommand{\revise}{\textcolor{black}}

\definecolor{red}{rgb}{1,0.0,0.0}

\definecolor{gray}{rgb}{0.3,0.3,0.3}
\newcommand{\code}[1]{\textcolor{gray}{\texttt{#1}}}

\newcommand{\ceil}[1]{\left\lceil #1 \right\rceil}

\begin{document}

\title{\name{}: A Constructive Interference-Assisted Reliable and Energy-Efficient LoRa Indoor Network}

\author{Mahbubur Rahman and Abusayeed Saifullah
\thanks{Mahbubur Rahman is with the City University of New York, and Abusayeed Saifullah is with Wayne State University. This manuscript is an extended version of a conference paper published in ACM/IEEE IoTDI '23~\cite{rahman2023boosting}.}}



\maketitle

\begin{abstract}
LoRa (Long Range) is a promising communication technology for enabling the next-generation indoor Internet of Things applications. Very few studies, however, have analyzed its performance indoors.
Besides, these indoor studies investigate mostly the RSSI (received signal strength indicator) and SNR (signal-to-noise ratio) of the received packets at the gateway, which, as we show, may not unfold the poor performance of LoRa and its MAC (medium access control) protocol -- LoRaWAN -- indoors in terms of reliability and energy-efficiency.
\revise{In this paper, we extensively evaluate the performance of LoRaWAN indoors and then use the key insights to boost its reliability and energy-efficiency by proposing {\em \name{} (\bf LoRa} {\bf I}ndoor {\bf N}etwork), a new link-layer protocol that can be effectively used for indoor deployments.
The approach to boosting the reliability and energy-efficiency in LoRaIN is underpinned by enabling constructive interference with specific timing requirements analyzed both empirically and mathematically for different pairs of channel bandwidth and spreading factor and relaying precious acknowledgments to the end-devices with the assistance of several booster nodes.}
The booster nodes do not need any special capability and can be a subset of the LoRa end-devices. 
To the best our knowledge, \name{} is the first protocol for boosting reliability and energy-efficiency in indoor LoRa networks. 
We evaluate its performance in an indoor testbed consisting of one LoRaWAN gateway and 20 end-devices. 
Our extensive evaluation shows that when 15\% of the end-devices operate as booster nodes, the reliability at the gateway increases from 62\% to 95\%, and the end-devices are approximately 2.5x energy-efficient.
\end{abstract}


\section{Introduction}\label{sec:introduction}

The next-generation Internet of Things (IoT) envisions reliable, energy-efficient, and scalable {\em indoor} applications through wireless connectivity between living things, machines, and sensors. These applications may include {\em continuous} Ammonia monitoring for the animals in laboratories or barns~\cite{jia2018continuous}, indoor localization~\cite{manzoni2019indoor, henriksson2016indoor}, industrial environment monitoring~\cite{haxhibeqiri2017lora}, patient monitoring~\cite{mdhaffar2017iot, hadwen2017energy}, and smart homes~\cite{lora_shome, varsier2017capacity}.
While low-power wide-area network (LPWAN) such as LoRa (Long Range) is adopted mostly outdoors~\cite{overview1, overview2, overview3, lpwan_survey1, ws_survey}, its ubiquity, popularity, low-cost, and scalability have made it a promising technology for indoors as well. Additionally, LoRa may be adopted indoors in contrast to the traditional WiFi or Bluetooth technology because of the following. (i) There are typically many users in the WiFi/Bluetooth band and relatively few users/applications in the subGHz band used by LoRa.
(ii) Being lower frequency, the LoRa band can better penetrate obstacles/walls than WiFi or Bluetooth. (iii)  WiFi is not well suited for very low traffic or small packet sizes, which are the characteristics of LoRa applications. Even though Bluetooth uses low traffic and small packets, its range is overly short and has lower penetration capability than the LoRa band.
Another motivating example is that LoRa and Comcast are soon to be conjoined in Comcast set-top boxes with the hope of proliferating smart home applications through macro, micro, and femto base stations/gateways~\cite{comlora, balanuta2020cloud}.

A LoRa system may involve one (or multiple) gateway(s) and numerous nodes (i.e., sensors) connected in a star (or star-of-stars) topology~\cite{lpwan_survey1}. To achieve reliability and different datarates, LoRa may employ different channel bandwidths (\code{BW}s) such as 125, 250, and 500kHz, spreading factors (\code{SF}s) such as 7, 8, 9, 10, 11, and 12, and different coding rates (\code{CR}s) such as $\frac{4}{5}$, $\frac{4}{6}$, $\frac{4}{7}$, and $\frac{4}{8}$ with a maximum transmission (Tx) power of 14dBm.
Despite its promises, only some studies have explored LoRa's potential in indoor environments~\cite{hadwen2017energy, mdhaffar2017iot, haxhibeqiri2017lora, henriksson2016indoor, manzoni2019indoor, gregora2016indoor, xu2019measurement}, which show that it is not well-suited for the indoor applications. 
Also, these studies analyzed mostly the RSSI (received signal strength indicator) and SNR (signal-to-noise ratio) of the received packets at a gateway under the LoRaWAN (LoRa wide-area network) MAC (media access control) protocol, which may not completely unfold the application scopes where reliability and energy-efficiency are very critical (e.g., in long-term monitoring applications). 
Furthermore, none of these works has focused on improving the communication reliability and energy-efficiency in an indoor LoRa network.

In this paper, we first evaluate the performance of LoRaWAN through experiments in an indoor environment, focusing primarily on the reliability and energy-efficiency at both the gateway and end-devices (i.e., nodes). 
In this experiment, we deploy 15 LoRa nodes and one gateway (capable of listening to 8 different channels simultaneously). We run this experiment for a week and find that the reliability at the LoRaWAN gateway can be as low as 62\%. Also, each node makes on average 5.2 transmission attempts to successfully deliver one packet to the gateway.
Such low reliability at the gateway despite the high number of Tx attempts per packet is ill-suited for critical indoor applications. Specifically, LoRaWAN performs worse under moderate to heavy network traffics due to
(1) the shadowing effect, higher path loss compared to outdoor, and interference between coexisting LoRa nodes and other devices operating in the same frequency band and 
(2) the LoRaWAN network server, which controls the gateway and acknowledges only the first reception of a packet by a node and never retransmits the acknowledgment (ACK) for any duplicate receptions to avoid the replay attack in the network~\cite{lwan_spec103}.
Consequently, if a node misses that ACK, its subsequent Tx attempts are wasted.

To boost the reliability and energy-efficiency indoors, we leverage the key insights of our initial experiments on LoRaWAN and propose {\em \name{}} ({\bf LoRa} {\bf I}ndoor {\bf N}etwork), a novel link-layer protocol for LoRa. Our approach to boosting the reliability and energy-efficiency in LoRaIN is underpinned by {\em creating constructive interference} at the gateway and {\em relaying precious ACKs} to the nodes, respectively, with the assistance of {\em booster nodes}. 
The booster nodes (or simply boosters) may be a subset of the LoRa nodes, which perform the following. (1) For an ongoing Tx, the boosters may listen to the packet and create constructive interference (hence improving the RSSI) such that it may be decodable by the gateway. (2) They may listen to a one-shot ACK and relay to the node that misses it, thereby stopping the subsequent redundant Tx attempts of a packet.
Note that boosting a signal far away may not result in a successful constructive interference due to the temporal displacement between a node and boosters, pathloss, and shadowing effect. The boosting in \name{} may thus be well-suited indoors only.

There are, however, a number of challenges to ensure the effective use of the boosters.
A booster must send an identical packet both at the same time and on the same channel to create a constructive interference to a packet of a node. Otherwise, this may lead to a two-packet collision scenario. Also,
if the booster fails to synchronize quickly with the packet, the node will suffer from high energy consumption due to many Tx attempts of the same packet. Moreover, the booster must ensure that it creates a constructive interference only if the node misses the one-shot ACK for the packet.
While relaying an ACK, a booster must also synchronize with both the gateway and the node that expects it. ACK relaying should be fast to avoid energy waste at the nodes. Additionally, a booster must not relay a duplicate ACK to any node that has already received it. In this paper, we address the above challenges and make the following key contributions.
\begin{itemize}
	\item To create constructive interference, a booster receives and synchronizes to the next Tx attempt of a packet 
	using the {\em carrier activity detection} (CAD) feature of the LoRa chip (e.g., SX1276) and the {\em receiving time window} as well as an {\em unused octet} of the LoRaWAN frame, respectively. To the best of our knowledge, this is the first attempt to create constructive interference in LoRa.

	\item A booster synchronizes (in both time and frequency) with both the gateway and a corresponding node to receive and relay an ACK using the node's {\em receiving time window}. Additionally, the booster compares the information on two customized octets of the LoRaWAN frame to suppress the duplicate ACKs. 

	\item We derive the maximum allowable temporal displacement between two LoRa transmitters in Matlab simulations for a successful constructive interference for any pairs of \code{BW} and \code{SF}. As an example, our Matlab simulations show that for a \code{BW} of 125kHz and \code{SF} of 10, the maximum allowable temporal displacement between two LoRa transmitters may not exceed 6.8243$\mu$s, which is considerably less than the corresponding chip duration ($\frac{1}{\code{BW}}$ = $\frac{1}{125000}$ = 8$\mu$s). \revise{Additionally, we provide an intuitive mathematical analysis for the timing requirements of constructive interference in \name{}, which is also consistent with our empirical analysis.} To the best of our knowledge, this is the first attempt to derive the maximum allowable temporal displacement between two LoRa transmitters for a successful constructive interference.

	\item We implement \name{} on 20 Dragino LoRa Hat nodes, each running on a Raspberry Pi, and one RAK2245 Pi Hat LoRaWAN gateway. We customize the LMIC 1.6 LoRaWAN code base to facilitate communications between the gateway, boosters, and other nodes.
	We then deploy these 20 nodes and the gateway in an indoor area of approximately 600ft$^2$. Our one-week-long experimental results show that when 15\% of the nodes act as boosters, the reliability in LoRa increases from 62\% to 95\%, and each node consumes $\approx$2.5x less energy, thus demonstrating the feasibility of \name{} with the commercial off-the-shelf devices.
\end{itemize} 



In the rest of the paper, Section~\ref{sec:related} overviews the related work. Section~\ref{sec:overview} overviews LoRa and its MAC protocol. \revise{Sections~\ref{sec:sysmodel},~\ref{sec:rationale}, and~\ref{sec:technical} describe the system model, design rationale, and the detailed design of \name{}, including both the empirical and mathematical timing analysis for constructive interference, respectively.} Section~\ref{sec:experiment} presents the implementation and experimental evaluation of \name{}.
Finally, Section~\ref{sec:conclusion} concludes our paper.




\section{An Overview of LoRa and LoRaWAN}\label{sec:overview}
In this section, we provide an overview of the LoRa physical layer, especially focusing on the encoding and decoding process of its waveform mathematically. Additionally, we provide an overview of LoRa’s MAC protocol, called LoRaWAN.
\subsection{LoRa Physical Layer}\label{sec:lora-phy}
\noindent{\bf PHY Overview.}
The LoRa PHY layer implements a {\em chirp spread spectrum (CSS)} modulation, where it encodes data using a linear frequency variation in a channel over time~\cite{peng2018plora}. To encode 0's (or chirp ``0") and 1's (or chirp ``1"), it differs the initial frequency in the chirps (a.k.a. symbols). In demodulation, a LoRa receiver multiplies an incoming chirp with a down-chirp 
and then applies a Fast Fourier Transformation (FFT). The FFT leads to a peak in a frequency bin, revealing the delay of the received chirp. The LoRa receiver decodes the chirp by tracking the location of that frequency bin. 
To make it more robust, a LoRa transmitter may use different \code{CR}s such as $\frac{4}{5}$, $\frac{4}{6}$, $\frac{4}{7}$, or $\frac{4}{8}$. 
To control the number of bits per chirp, it may also use different \code{SF}s between 7 and 12. Additionally, the LoRa PHY may choose different channel bandwidths (e.g., \code{BW}s such as 125, 250, or 500kHz).
As the chirps fully utilize the channel bandwidth \code{BW}, the LoRa PHY layer becomes resilient (to some extent) to the Doppler and multi-path effects and channel noise.

\noindent{\bf PHY Encoding.}
\revise{The CSS signals are constructed using complex sinusoid with linear frequency variation over frequency range $[-\frac{\code{BW}}{2}, \frac{\code{BW}}{2}]$ and time range $[0, T]$, where $T$ is the LoRa symbol duration. The basic LoRa signals are up-chirps and down-chirps whose frequencies change linearly from $-\frac{\code{BW}}{2}$ to $\frac{\code{BW}}{2}$ and $\frac{\code{BW}}{2}$ to $-\frac{\code{BW}}{2}$, respectively. Owing to the properties of discrete-time complex signal processing, the Nyquist sampling period $T_s = \frac{1}{\code{BW}}$ is used at the transmitter of a LoRa signal. For a spreading factor of \code{SF}, each LoRa symbol consists of \code{SF} bits, resulting in an $M$-ary modulation scheme with $ M = 2^{\code{SF}}$. We thus have $T_s = \frac{1}{\code{BW}} = \frac{T}{M}$ and a LoRa symbol consisting of $M$ chips or samples (assuming one sample per chip). For encoding, a LoRa symbol $\alpha \in \{0, 1, \cdots, M-1\}$ is mapped to an up-chirp that is shifted in time by a period of $\tau_{\alpha} = \alpha T_s$. Such a shift in time refers to a linear variation of the frequency by $\frac{\alpha \code{BW}}{M} = \frac{\alpha}{MT_s} = \frac{\alpha}{T}$~\cite{demeslay2021theoretical}. A modulo operation is also applied to ensure that the linear variation in frequency stays in the interval $[-\frac{\code{BW}}{2}, \frac{\code{BW}}{2}]$. 
Finally, a mathematical expression for the discrete-time LoRa symbol wave-form $s_{\alpha}[n]$, sampled at time $t = nT_s$, may be expressed as follows~\cite{chiani2019lora}.
\begin{equation}
\label{eqn:lora_symbol}
s_{\alpha}[n] = s_{\alpha}(t)|_{t = nT_s} = e^{j2\pi n(\frac{\alpha}{M} - \frac{1}{2} + \frac{n}{2M})}; n = 0, 1, \cdots, M-1.
\end{equation}
The Equation (\ref{eqn:lora_symbol}) above represents an up-chirp for $\alpha = 0$ and may thus be denoted by $s_0[n]$. A down-chirp, on the other hand, is the complex conjugate of $s_0[n]$ and denoted by $s_0^*[n]$.}

\noindent{\bf PHY Decoding.}
\revise{The demodulation process of the LoRa symbols includes multiplying an incoming symbol by a down-chirp, followed by applying an FFT on the multiplication result. The FFT leads to a peak in a frequency bin, revealing the delay of the received symbol. Mathematically, the overall process may be derived as follows, which is based on the {\em maximum likelihood} detection scheme~\cite{vangelista2017frequency}.
In an additive white Gaussian noise (AWGN) channel with a complex noise $w[n]$ having a \code{zero} mean and $\sigma^2 = E[|w[n]|^2]$ variance, the received LoRa symbol $r[n]$ may be represented as
\begin{equation}
\label{eqn:lora_rx}
r[n] = s_{\alpha}[n] + w[n].
\end{equation}
To this extent, the goal of the maximum likelihood detector at the receiver is to choose a symbol index $\widehat{\alpha}$ that maximizes the multiplication $\langle r[n], s_m[n] \rangle$ for $m \in \{0, 1, \cdots, M-1\}$. The overall mathematical process at the receiver may be defined as follows~\cite{demeslay2021theoretical}.
\begin{align}
	\nonumber \langle r[n], s_m[n] \rangle &= \sum_{n=0}^{M-1} r[n]s_m^*[n] \\
                     &= \sum_{n=0}^{M-1} r[n]s_0^*[n]e^{-j2\pi \frac{m}{M}n} \label{eqn:lora_dmod}\\
               \nonumber        &= \sum_{n=0}^{M-1} \tilde{r}[n] = \tilde{R}[m]\
\end{align}
where $\tilde{r}[n] = r[n]s_0^*[n]$. In the above derivation, the complex conjugate $s_m^*[n]$ of $s_m[n]$ is required since that would lead towards the maximum degree of similarities between $r[n]$ and $s_m[n]$~\cite{chaudhari2018wireless}. A closer look at Equation (\ref{eqn:lora_dmod}) also reveals that the received symbol waveform $r[n]$ is multiplied by the down-chirp $S_0^*[n]$ (this process is known as de-chirping in the literature) and $\tilde{R}[m]$ is obtained through discrete Fourier transform (e.g., FFT) of $\tilde{r}[n]$.
The above process combines all the symbol signal energy into a unique FFT frequency bin and may be retrieved by taking the magnitude of $\tilde{R}[m]$.
Afterwards, the detected symbol
\begin{equation}
\widehat{\alpha} = \arg \max_{m} |\tilde{R}[m]|^2. \nonumber
\end{equation}}

\begin{figure}[!htbp]
\centering \vspace{-0.2in}
\includegraphics[width=0.35\textwidth]{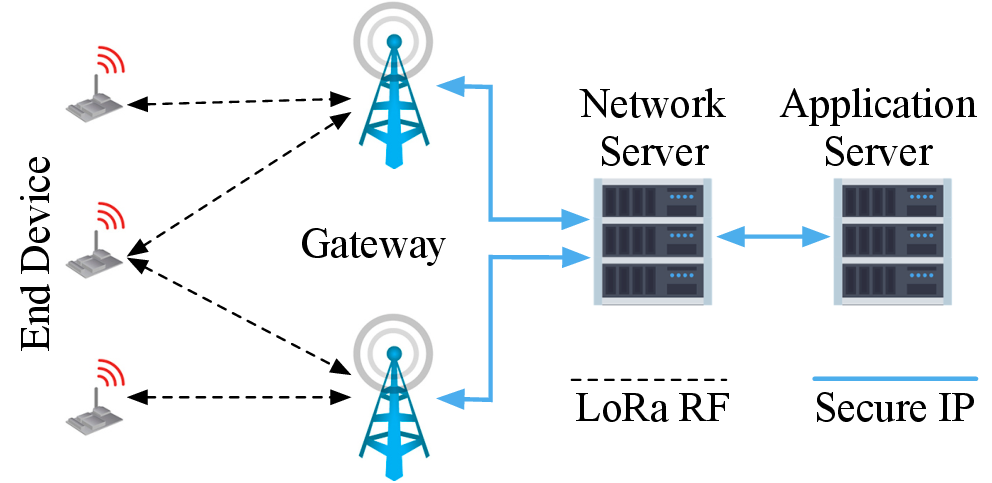}\vspace{-0.15in}
\caption{A typical LoRaWAN architecture.} \vspace{-0.15in}
\label{fig:lorawan} \vspace{-0.05in}
\end{figure}
\subsection{LoRaWAN Architecture and Basics}
As shown in Figure~\ref{fig:lorawan}, a LoRaWAN network consists of end-devices (i.e., nodes/sensors), one or more gateways, a network server, and one or more application servers. The LoRaWAN frequency band is divided into two parts: multiple {\em uplink} and multiple {\em downlink} channels. The nodes send data to the gateways over the uplink channels. The gateways then pass the data to the network server. The network server deduplicates (if necessary) and sends the data to the application server, as necessary. On the other hand, the network server may send messages (e.g., for management or on behalf of the application server) through the gateways. The gateways communicate with the nodes using the downlink channels. LoRaWAN categorizes the nodes in three classes (class-A, class-B, and class-C) based on when they want to receive downlink messages. These classes directly determine the energy-efficiency of the nodes. In the following, we briefly discuss these classes.

\subsubsection{LoRaWAN Classes}\label{sec:class-a}
In LoRaWAN, all the nodes are required to support {\em class-A} (``Aloha"). A node may spend most of the time in sleep mode. The node can communicate with the network server (through a gateway) anytime it wants. After sending an uplink message, it may listen for a message from the network server one or two seconds before going back to sleep. This is the most energy-efficient class of LoRaWAN.
In {\em class-B} mode, a node wakes up and opens receive windows to listen for downlink messages according to a configurable but network-defined schedule. A periodic beacon signal from the network server allows the class-B nodes to synchronize their internal clocks with the network server. 
The LoRaWAN {\em class-C} (``Continuous") nodes never go to sleep. They constantly listen for downlink messages from the network server, except when they have their own data to transmit. As a result, they consume the most energy across all the classes.

\subsubsection{Unconfirmed and Confirmed Messaging}\label{sec:con-uncon-msgs}
A message from a node to the network server and vice versa may be {\em confirmed} or {\em unconfirmed}. In confirmed messaging, the sender requests an ACK from the receiver. When a node sends a confirmed message to the network server, it makes up to 8 Tx attempts until it gets an ACK.
In unconfirmed messaging, a sender does not request an ACK from the receiver.

\section{\name{} System Model}\label{sec:sysmodel}
We consider the indoor applications that require high reliability and energy-efficiency at the nodes. \name{} may have one or multiple gateways and numerous nodes in the network. 
We, however, evaluate \name{}'s performance with one gateway and a subset of the nodes working as the boosters. This setup is similar to the idea of having one Wi-Fi access point and a few range extenders (if needed) to improve the WiFi network performance in a home or indoor scenario. Additionally, using multiple gateways instead of boosters may be cost prohibitive. A commercial gateway may cost, on average, US\$250 (8-channel) to US\$2,494 (16-channel)~\cite{gprices}. 
Having boosters instead of multiple gateways is thus an economic and favorable solution indoors since they are a subset of the nodes and incur no additional cost.
In \name{}, the gateway is wall-powered, while the nodes (including the boosters) can be either wall-powered or battery-powered (which gives much freedom of installation and avoids wiring cost and complexity in the smart building use cases~\cite{batterypowered}). Wall-powered or battery-powered, 
it is beneficial to avoid or nullify interference in any network, which may be caused by redundant retransmissions by the nodes (as explained in Section~\ref{sec:introduction}). \revise{In \name{}, we achieve the above while providing energy-efficiency in the nodes using boosters that use the {\em ultra-low-power} CAD feature to participate in the boosting activities. We analyze the energy overhead (which is crucial if battery-powered) in the boosters in Section~\ref{sec:booster_energy}, which shows their ability to improve (by 2.5x) the overall energy efficiency in the network (including theirs) while consuming $<$1 mJ energy per bit in boosting.}

In \name{}, we adopt the class-A mode of LoRaWAN in the nodes because of its energy-efficiency. To ensure the reliability in data transfer, we adopt the confirmed uplink messaging of LoRaWAN. Note that a node makes up to 8 Txs to get an ACK in confirmed uplink messaging. The network server acknowledges (via the gateway) only the first received Tx of a packet by a node and never retransmits it. Similar to LoRaWAN, we do not allow the gateway to send multiple ACKs for multiple Tx attempts of the same packet. The reasons for this are as follows. (1) If the first ACK is not received by the node, it is highly likely that the node will not receive the following ACKs as well. This may be because of the bad link quality between the gateway and the node. (2) Allowing multiple ACKs may result in a replay attack in the network.
In the rest of the paper, we denote the messages from the network server (via the gateway) to the nodes simply as the messages from the gateway to nodes. Also, we denote the messages from the nodes to the network server (via the gateway) as the messages from the nodes to the gateway.

\begin{figure}[!htbp]
\centering  
\includegraphics[width=0.35\textwidth]{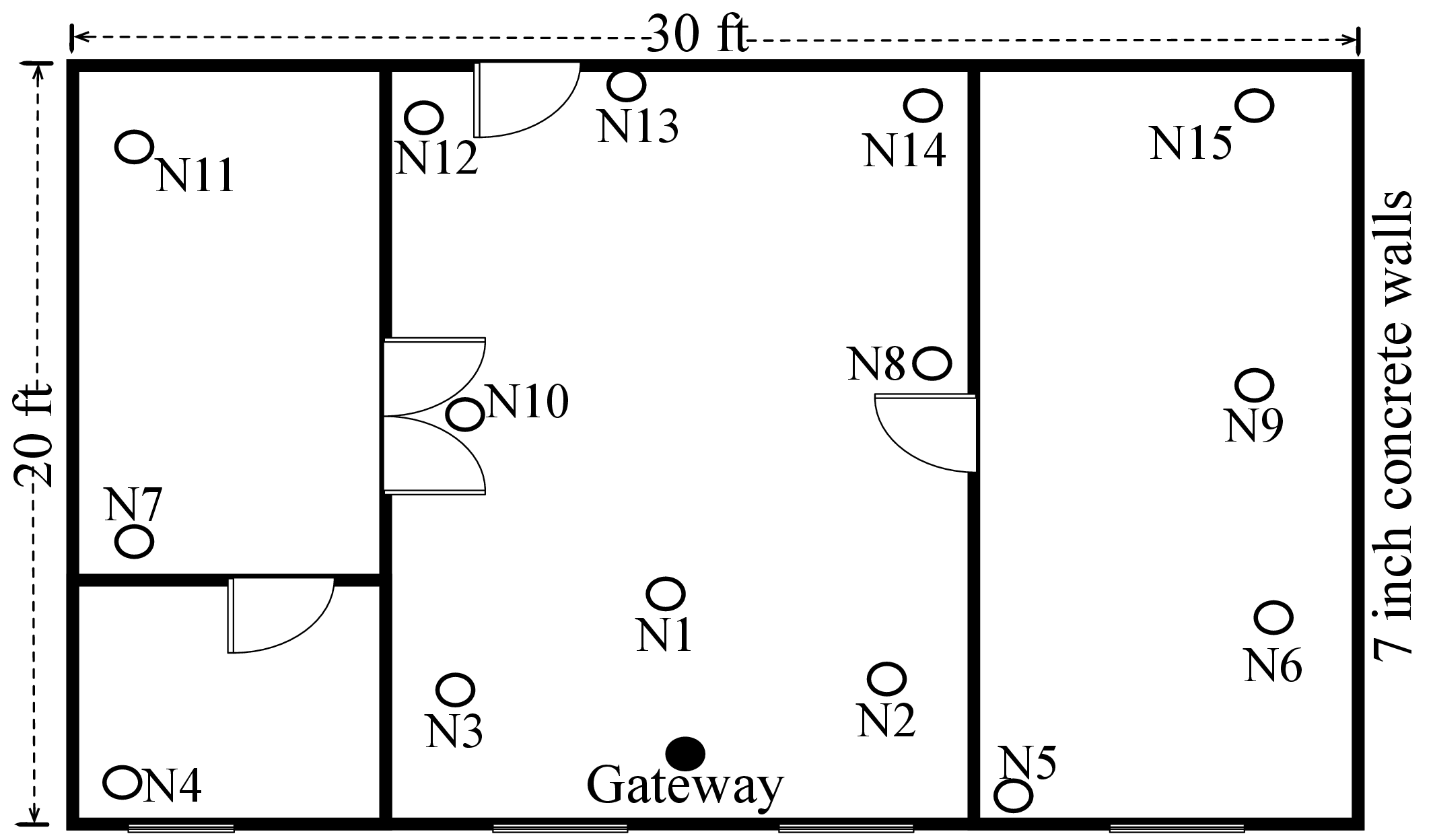}  
\caption{Locations of the gateway and the nodes.}  
\label{fig:indoor} 
\end{figure}
\section{\name{} Design Rationale}\label{sec:rationale}
We now analyze the performance of LoRaWAN indoors. Specifically, we analyze the reliability and energy requirements for both the gateway and the nodes. We then use these analyses to make design decisions in \name{}. \revise{The experimental dataset is available online~\cite{loraindata}. In the following, we first explain our experimental setup.}

\begin{figure}[!htbp] 
    \centering 
      \subfigure[PRR at the gateway\label{fig:prr_gw1}]{ 
    \includegraphics[width=0.28\textwidth]{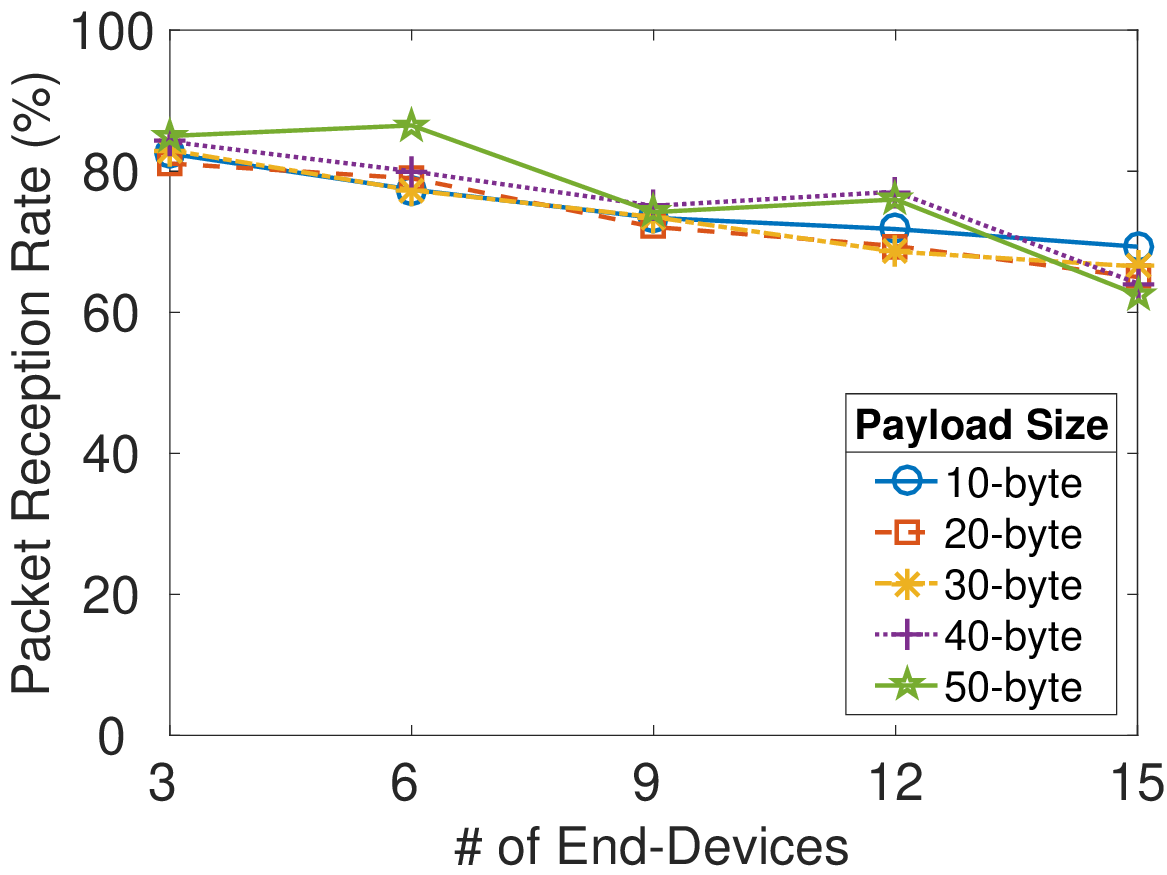}
      }\vfill 
      \subfigure[PDR at the nodes\label{fig:pdr_nodes}]{
        \includegraphics[width=.28\textwidth]{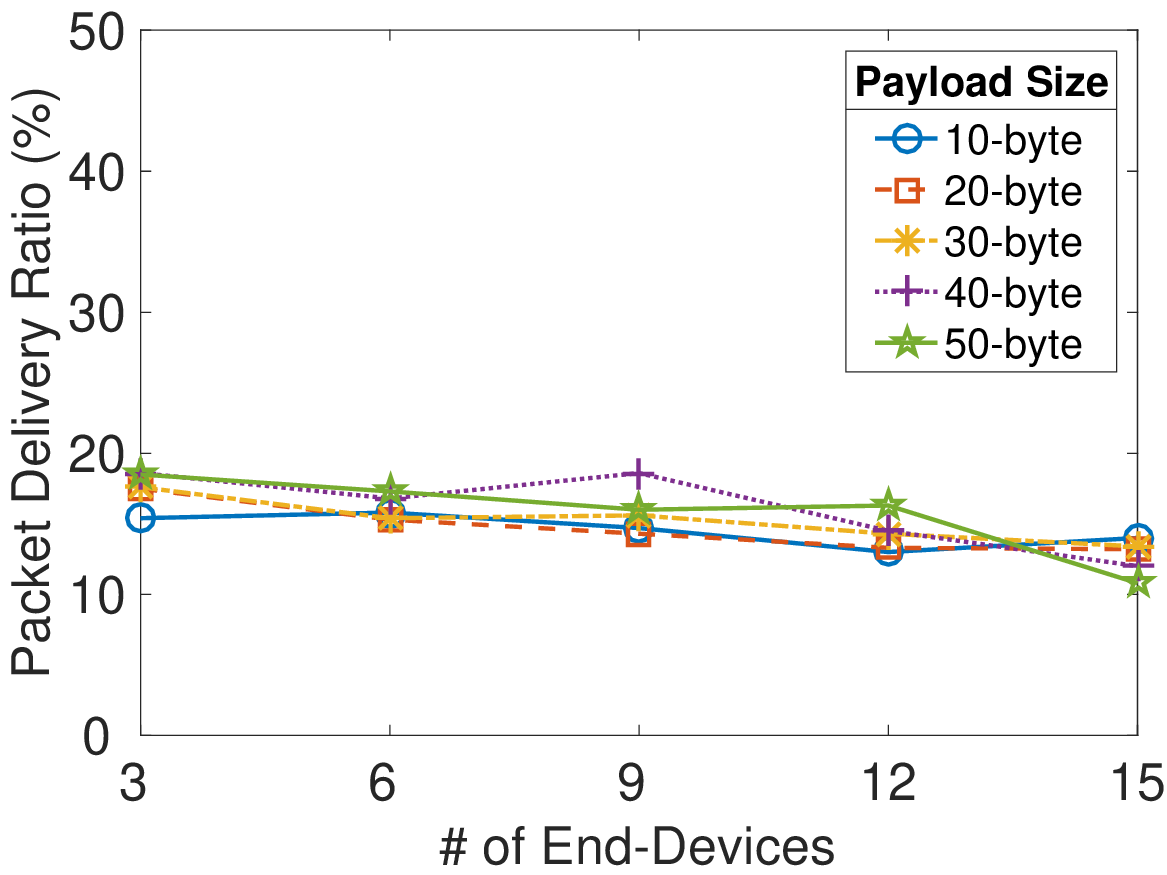}
      } 
    \caption{Reliability analysis of LoRaWAN in indoor.} \vspace{-0.15in}
    \label{fig:lora_reliability}
\end{figure}
\subsection{Experimental Setup}\label{sec:rationale_setup}
\revise{We deploy a LoRaWAN network in an indoor area of approximately 600ft$^2$, as shown in Figure~\ref{fig:indoor}. We use one LoRaWAN gateway (represented by a dark circle in the figure) and 15 nodes ( represented by empty circles and labeled N1--N15, respectively, in the figure). In this setup, the gateway runs a ChirpStack LoRaWAN network server~\cite{cstack} locally and the nodes operate in LoRaWAN class-A mode. Our gateway is capable of receiving in eight 125kHz uplink channels. The gateway also uses eight 500kHz downlink channels to send the ACKs. The geographic location of our network does not have any duty cycle requirements in the channels and does not allow a \code{SF} of 12 as well. The {\em adaptive data rate} (ADR) feature of LoRaWAN is enabled at both the network server and the nodes. ADR adapts the \code{SF} and \code{CR} dynamically to improve the LoRaWAN signal/packet receptions. In our experiments, we find that
the \code{SF} varies between 7 and 10 while the \code{CR} is fixed at $\frac{4}{5}$. Also, the average RSSI and SNR at the gateway are approximately -44.7dBm and 9.5dB, respectively, when we transmit packets with a Tx power of 14dBm. Unless stated otherwise, these are our default experimental setup for the \name{} design rationale.}

\subsection{Reliability Analysis of LoRaWAN}\label{sec:analysis_reliability}
In this section, we analyze the reliability at the LoRaWAN gateway and various number of nodes for confirmed uplink communication. With the setup in Section~\ref{sec:rationale_setup}, each node sends 100 confirmed uplink packets with an inter-packet interval of 1 minute. This interval is common in many indoor applications such as Ammonia monitoring for barn animals~\cite{jia2018continuous} and patient monitoring in hospitals. We send packets from 3 to 15 nodes with different payload lengths between 10 and 50 bytes. Figure~\ref{fig:lora_reliability} shows the reliability in the gateway and the nodes in the forms of {\em packet reception rate} (PRR) and {\em packet delivery ratio} (PDR), respectively. PRR at the gateway is defined as the ratio of the number of packets received at the gateway to the total number of packets sent by the nodes. On the other hand, PDR at the nodes is defined as the ratio of the number of ACKed packets to the number of total packets sent by the nodes. Since there is no idea of ACK in PRR, we use two different metrics (as detailed below) to evaluate the reliability at the gateway and nodes, respectively.

\subsubsection{Packet Reception Rate}\label{sec:prrlow}
As shown in Figure~\ref{fig:prr_gw1}, the PRR at the LoRaWAN gateway is approximately 82.5\% when 3 nodes transmit to the gateway with a payload of 10 bytes. As the number of nodes increases, the PRR at the gateway decreases significantly. For example, when 15 nodes transmit 10-byte payloads, the PRR goes down to approximately 69.3\%. Figure~\ref{fig:prr_gw1} also shows that this decreasing trend in the PRR is steady for all packet sizes. When 15 nodes transmit 50-byte packets, the PRR at the gateway is as low as 62\%. LoRa observes such low PRR at the gateway because of the interference due to severe multi-path and shadowing effects in indoor and packets (on the same channel) collisions, resulting in many packets being lost. Although the LoRa modulation allows the gateway to recover packets below the noise floor, the gateway may not be able to decode a packet residing within the above interference scenario. The reason is that the FFT at the gateway may not be able to distinguish between the up-chirps and down-chirps in the received signal because of the data availability (or unavailability) in the undesired (or desired) frequency bins (Section~\ref{sec:lora-phy}).

\begin{figure}[!htbp]
    \centering 
      \subfigure[Average Tx attempts per packet\label{fig:avg_attempts}]{
    \includegraphics[width=0.28\textwidth]{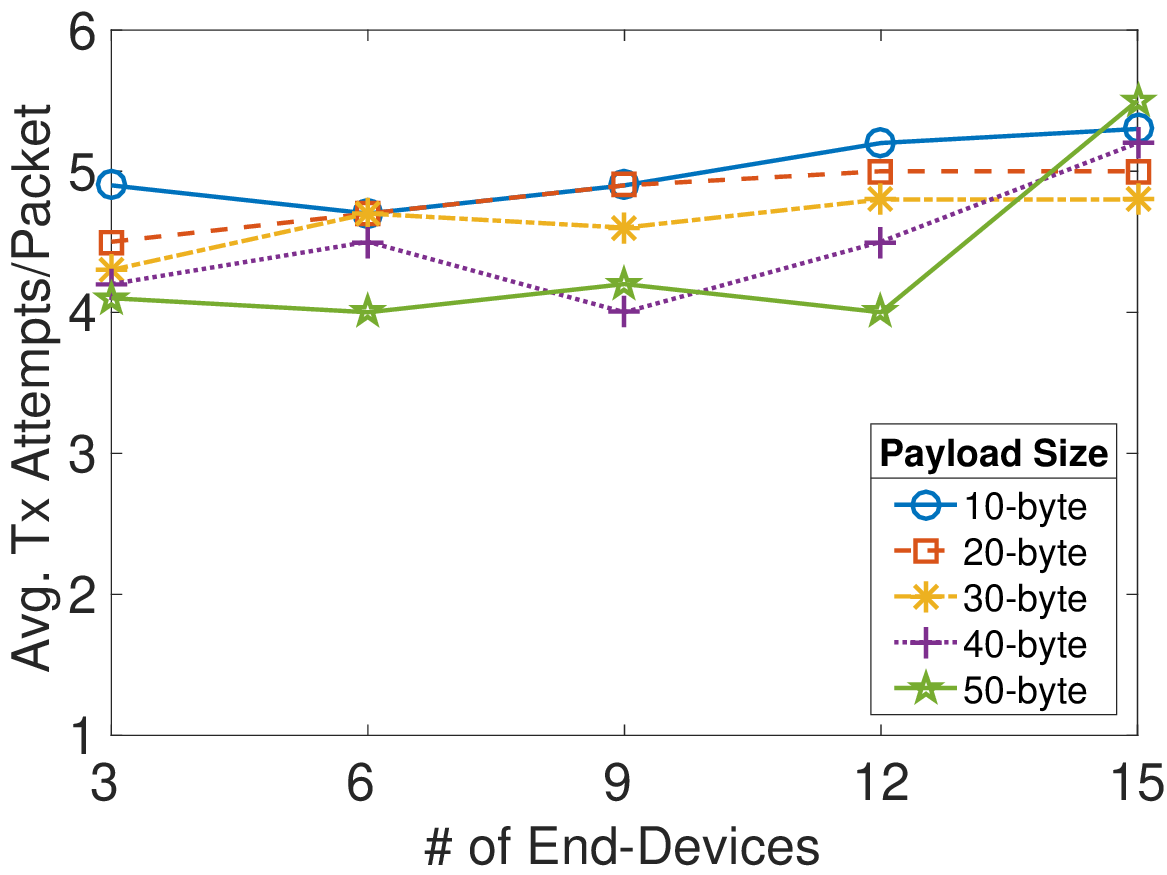}
      } \vfill 
      \subfigure[Lost ACK count at the nodes\label{fig:lack_nodes}]{
        \includegraphics[width=0.28\textwidth]{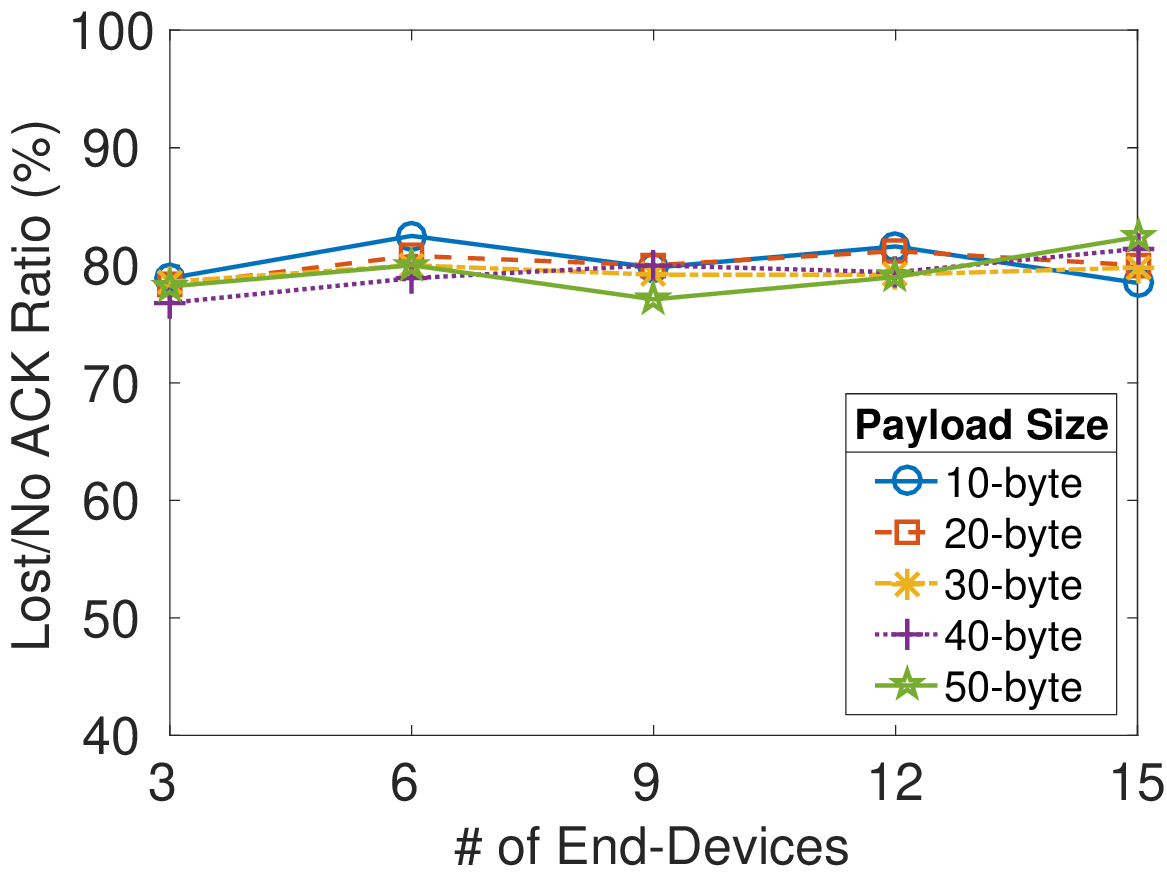} 
      } 
    \caption{Number of Tx attempts and ACKs in LoRaWAN.} 
    \label{fig:lora_attempts}
\end{figure}

\subsubsection{Packet Delivery Ratio}\label{sec:pdrlow}
As shown in Figure~\ref{fig:pdr_nodes}, when 3 nodes send 10-byte payloads, the average PDR is approximately 15.4\% at the nodes. Also, as the number of nodes increases, they observe even lower PDRs. For example, the average PDR at the nodes is approximately 14\% when 15 nodes send 10-byte payloads. Figure~\ref{fig:pdr_nodes} also shows that as we vary the payload size, the packet delivery ratios at the nodes follow a similar pattern. When 15 nodes transmit 50-byte payloads, the average PDR is approximately 10.8\%. The nodes observe such low PDRs due to the following two potential reasons. (1) The gateway cannot decode the received signals and thus does not send ACKs. (2) The ACK sent for a packet is not received/decoded at the corresponding node.

\subsubsection{Discussion}
Both PRR and PDR in this experiment are considerably very low, while the latter is much worse. This also means that a large number of correctly decoded packets at the gateway are redundantly retransmitted by the nodes. To this extent, we propose to boost the reliability of indoor LoRa by introducing \name{}. 




\subsection{Energy Requirement Analysis of LoRaWAN}\label{sec:analysis_energy}
We now analyze the energy requirements at the nodes. 
Equation (\ref{eqn:eattempts}) below estimates the relationship between the energy consumption and the Tx attempts for a packet.
\begin{equation}
\label{eqn:eattempts}
E_{\text{packet}} \approx N_{\text{attempt}} \times (E_{\text{air}} + E_{\text{Rx1}} + E_{\text{Rx2}})
\end{equation}
Here, $E_{\text{packet}}$ is an estimation of the packet's total energy consumption, $N_{\text{attempt}}$ is the total attempts for a packet, $E_{\text{air}}$ is the energy consumption for each Tx airtime, $E_{\text{Rx1}}$ is energy consumption in the first receive (Rx) window, and $E_{\text{Rx2}}$ is the energy consumption in the second Rx window. For simplicity, we do not include the energy consumption for the Tx--Rx radio switches and Rx-delays between the Tx window and Rx windows as their contributions are negligible.
In the following, we analyze the Tx attempts by the nodes with the setup explained in Section~\ref{sec:rationale_setup}.
\begin{figure}[!htbp]
    \centering 
      \subfigure[CDF of Tx attempts\label{fig:cdf_attempts}]{ 
        \includegraphics[width=.28\textwidth]{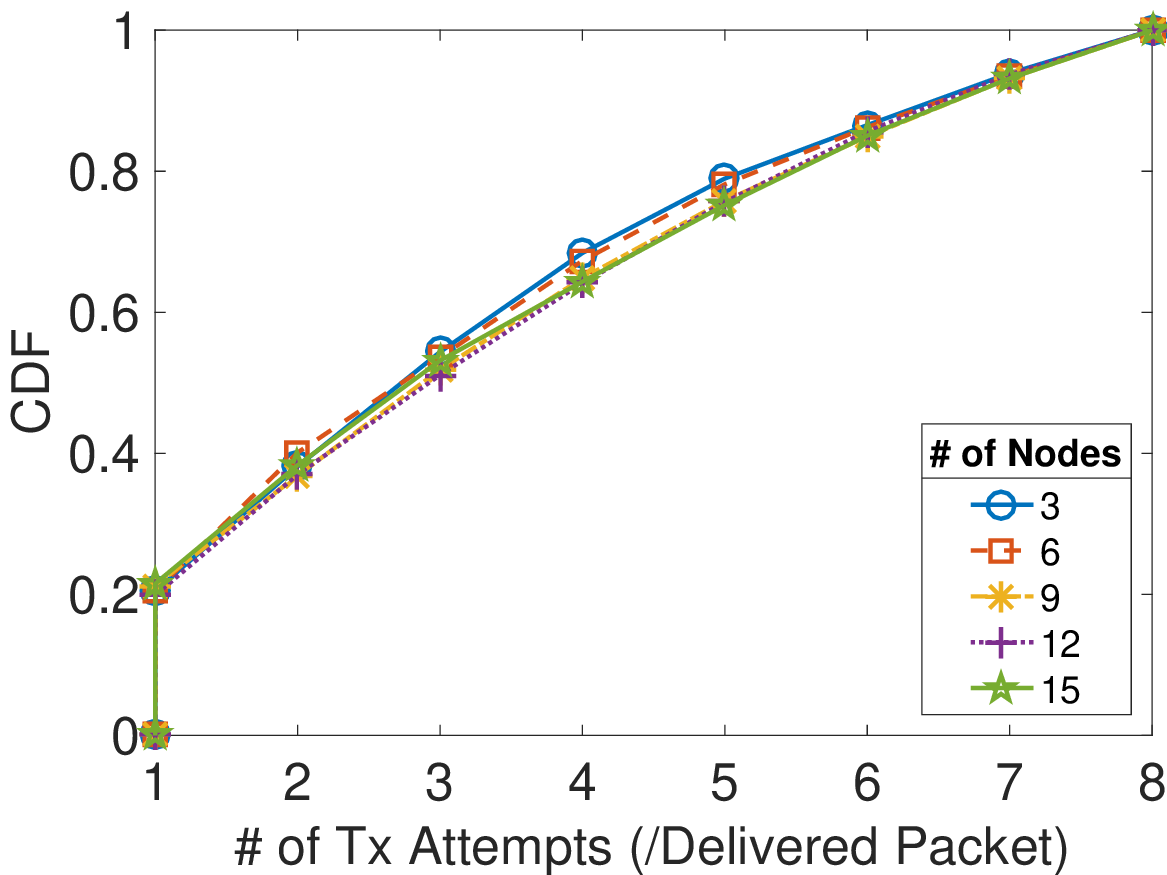}
      } \vfill 
      \subfigure[Counts of Tx attempts\label{fig:histo_attempts}]{
        \includegraphics[width=0.28 \textwidth]{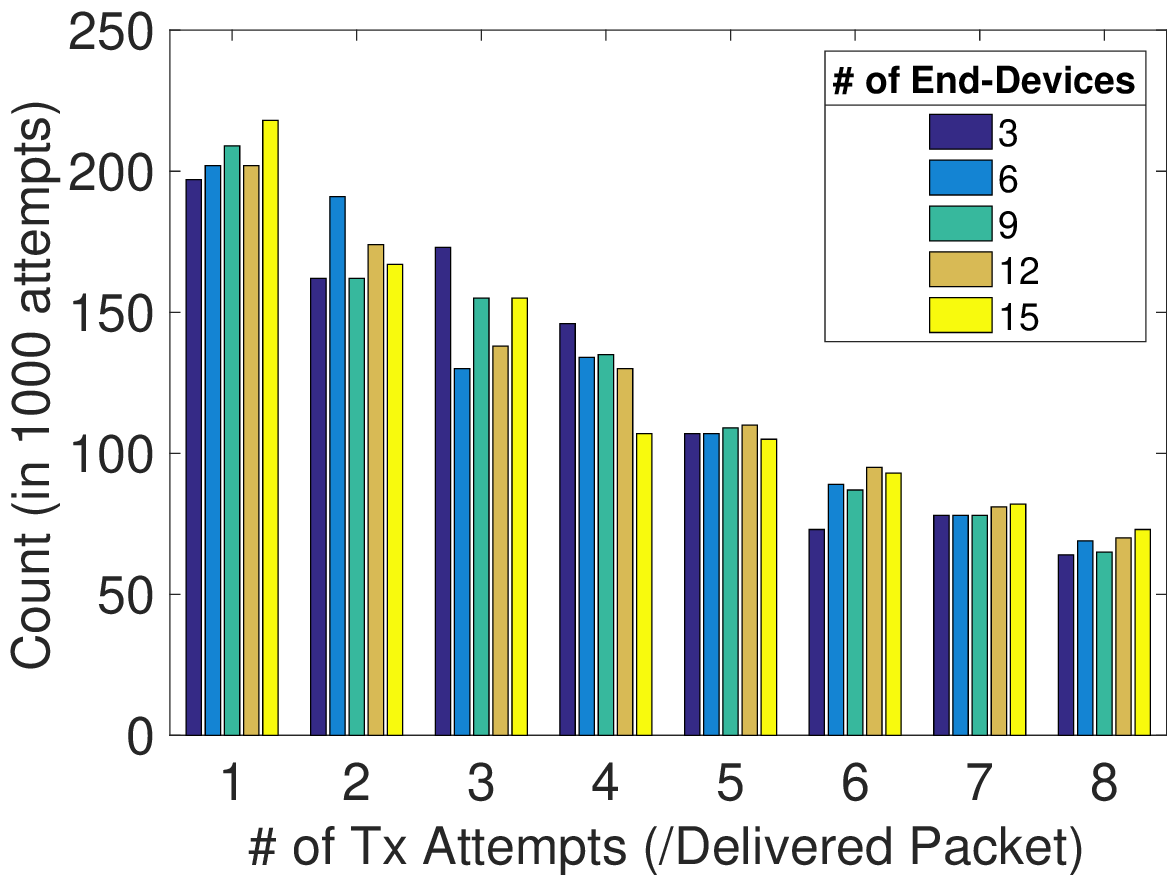} 
      } 
    \caption{Number of Tx attempts for the delivered packets.} 
    \label{fig:tx_attempts}
\end{figure}
\subsubsection{Considering All the Packets in Uplink}
Here, we analyze the number of Tx attempts while considering all the packets scheduled from the nodes to the gateway. 

\noindent{\bf All Transmission Attempts.}
Figure~\ref{fig:avg_attempts} shows the average number of Tx attempts per packet for various number of nodes. When 3 nodes send 100 packets each, each packet with a 10-byte payload, the average number of Tx attempts per packet is approximately 4.9. This figure also shows that as the number of nodes increases, the average number of Tx attempts per packet also increases. When 15 nodes transmit packets with 10-byte payloads, the average number of Tx attempts is approximately 5.3. This increasing trend in the average number of Tx attempts remains steady for the packets of different payload sizes. For example, the average number of Tx attempts per packet is approximately 4.1 compared to 5.5 when 3 nodes and 15 nodes send packets, each with 50-byte payloads, respectively. The main reasons for such high numbers of Tx attempts by the nodes are twofold. (1) The gateway sends an ACK but the corresponding node is unable to receive the ACK. (2) The gateway does not send ACK for the subsequent Tx attempts by a node for which an ACK has already been sent for a prior Tx  attempt. In the following, we further investigate the above two cases.

\noindent{\bf Lost/No Acknowledgment Count.}
In Figure~\ref{fig:lack_nodes}, we show the analysis on the number of ACKs for all the packets in the uplink. Here, for any number of nodes between 3 and 15 that send packets with payload sizes between 10 to 50 bytes, the aggregate numbers of lost (at the nodes) and unsent (by the LoRaWAN server) ACKs vary approximately between 77\% and 83\%. 
Such a poor performance due to the design choices of the LoRaWAN network server can be considered as a serious drawback for energy-constrained IoT nodes.
In fact, lost/no ACKs is the most critical reason for the large number of unnecessary Tx attempts by the nodes.

\noindent{\bf Discussion.}
Figure~\ref{fig:lora_attempts} shows that it is critical to boost the performance of LoRaWAN in terms of the number of Tx attempts by the nodes to deliver a packet. Similarly, LoRaWAN needs a robust ACK mechanism so that it may enable high PDRs and reduce the average number of Tx attempts per packet, thereby not wasting node's invaluable and limited energy budgets.

\subsubsection{Considering the Delivered Packets}

\revise{We now analyze the number of Tx attempts of the packets for which ACKs are received by the nodes (thus considered {\em delivered}). To unfold the energy requirements more closely, we consider two scenarios: all delivered packets and pairs of packets for which acknowledgments are received consecutively. These two scenarios will help us make design decisions in \name{} to boost the reliability and energy-efficiency.}

\noindent{\bf All Delivered Packets.}
Figure~\ref{fig:cdf_attempts} shows the cumulative distribution function (CDF) of the number of Tx attempts of all the packets for which ACKs are received by the nodes. When 3 to 15 nodes transmit with payloads of sizes between 10 and 50 bytes, approximately 20\% of the packets are delivered to the gateway through single Tx attempts, and approximately 40\% of the packets require 1 to 2 Tx attempts. The rest of the packets ( approximately 60\%) need up to 8 Tx attempts, which results in huge energy consumption at the nodes. Additionally, we take such 1000 packets with payload sizes varying between 10 and 50 bytes from the experiments and plot the numbers of Tx attempts by the nodes in Figure~\ref{fig:histo_attempts}. As shown in this figure, when 3 to 15 nodes transmit packets with payloads of sizes between 10 and 50 bytes, on average 650 out of 1000 packets need 3 or more Tx attempts.

\begin{figure}[!htbp]
    \centering 
      \subfigure[CDF of Tx attempts\label{fig:cdf_conse_attempts}]{
    \includegraphics[width=0.28\textwidth]{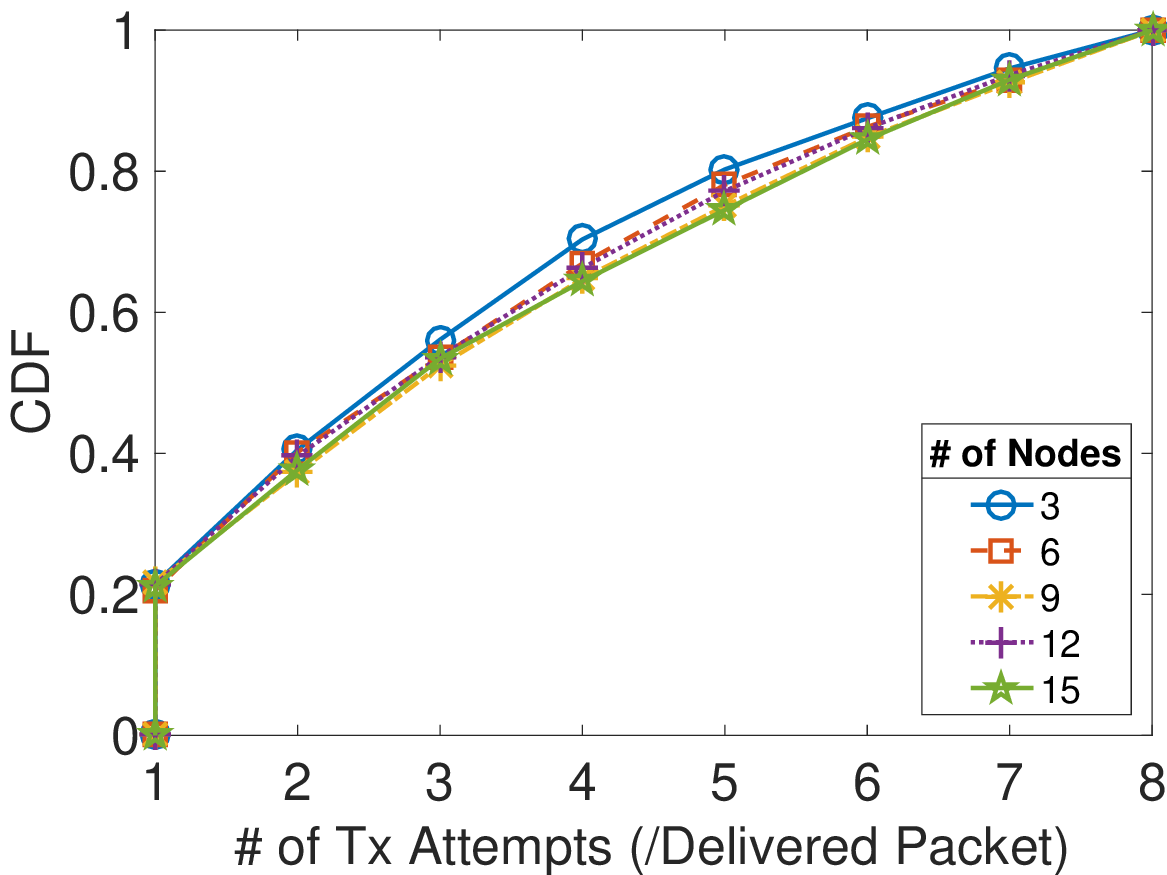}
      }\vfill 
      \subfigure[Counts of Tx attempts\label{fig:histo_conse_attempts}]{
        \includegraphics[width=.28\textwidth]{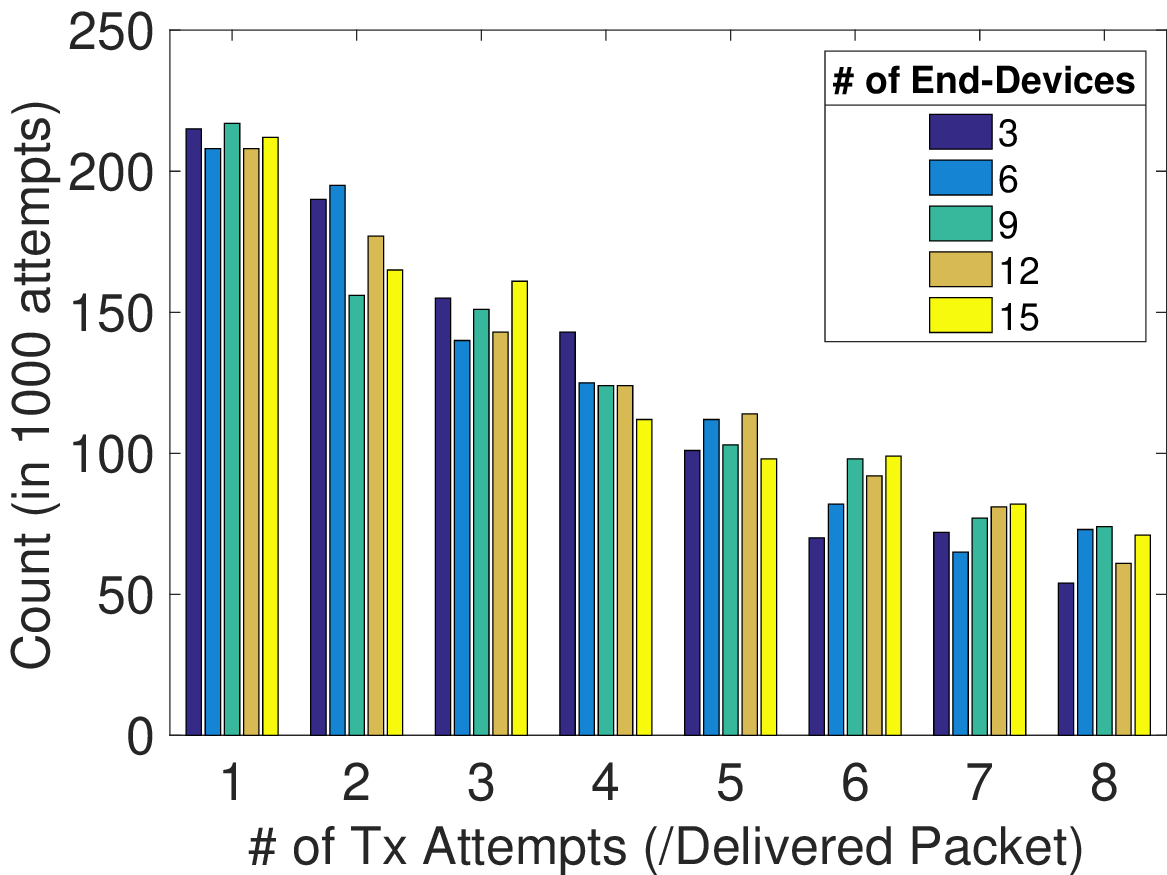}
      } 
    \caption{\revise{Number of Tx attempts for pairs of the delivered packets.}} \vspace{-0.2in}
    \label{fig:tx_consec_attempts}
\end{figure}

\noindent{\bf Pairs of Consecutively Delivered Packets.} 
\revise{Figure~\ref{fig:cdf_conse_attempts} shows the CDF of the numbers of Tx attempts of the packets for which ACKs are received consecutively at the nodes. As shown in this figure, even for the consecutively delivered pairs of packets, the numbers of Tx attempts by the nodes are almost identical to those analyzed in Figure~\ref{fig:cdf_attempts}. For example, when 3 to 15 nodes transmit packets with payload sizes between 10 and 50 bytes, approximately 60\% of the packets require up to 8 Tx attempts by the nodes. Similarly, analyzing 1000 of such packets (i.e., 500 pairs of consecutively delivered packets), we find that LoRaWAN still causes very high energy consumption at the nodes. As shown in Figure~\ref{fig:histo_conse_attempts}, when 3 to 15 nodes transmit packets with payload sizes between 10 and 50 bytes, on average 625 out of 1000 packets still need 3 or more delivery attempts by the nodes.}

\noindent{\bf Discussion.} 
\revise{As we analyze the number of transmission attempts for the delivered packets, we find that LoRaWAN gateway indeed misses a lot of packets, requiring the nodes to make subsequent transmission attempts that are necessary to deliver the packets. Overall, our analysis (as shown in Figures~\ref{fig:lora_attempts},~\ref{fig:tx_attempts}, and~\ref{fig:tx_consec_attempts}) reveals that the LoRaWAN network server performs poorly in indoor scenarios in terms of energy requirements at the nodes and reliability at both the gateway and nodes. Hence, it is important to boost the energy-efficiency in indoor LoRaWAN. To this extent, we detail the design of \name{} in the subsequent sections.}

\section{Design of \name{}}\label{sec:technical}

In this section, we discuss the detailed design of LoRaIN, including an overview and the protocols developed for boosting the reliability and energy efficiency of LoRa indoors.
\subsection{Design Principles}

\subsubsection{Booster Selection}
In \name{}, we boost the reliability and energy-efficiency by introducing boosters in the network. Boosters are a subset of the LoRa nodes in the network. These nodes may be selected during the network deployment phase or later. Due to the uncertain noise or interference characteristics in the indoor environments, the number of boosters may be decided dynamically by the LoRaWAN network/application server from the set of LoRa nodes (e.g., nodes that observe better energy efficiency or that are application specific). For this purpose, the network/application server may retain an editable/configurable list of these boosters. \revise{As needed, the gateway may also request or allocate one or multiple boosters for specific channels that observe too many packet losses.}
\begin{figure}[!htbp]
\centering 
\includegraphics[width=0.4\textwidth]{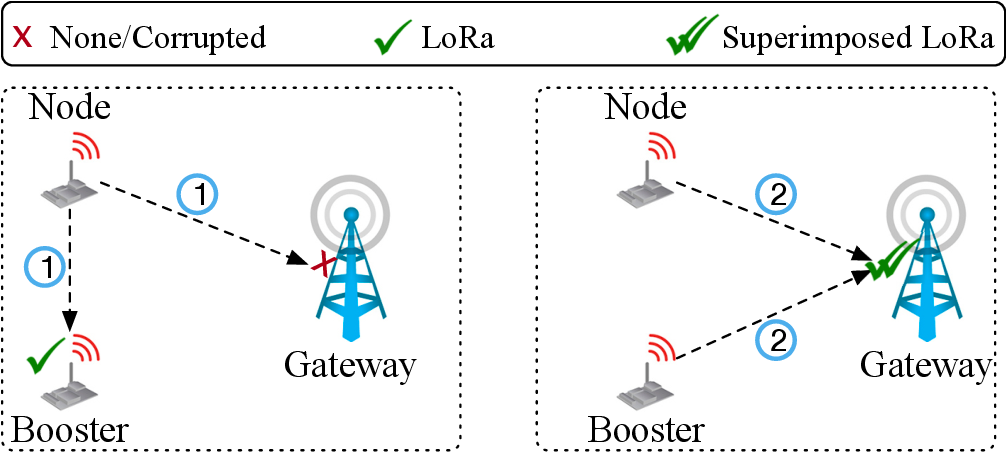} 
\caption{Steps (denoted by numbered circles) of boosting the reliability in \name{}. "x" represents no/corrupted LoRa signal/packet, "single tick" represents a valid LoRa signal/packet, and "double tick" represents a superimposition of multiple LoRa signals/packets.} 
\label{fig:lorainup} 
\end{figure}

\subsubsection{Boosting Reliability}
The LoRaWAN gateway greatly suffers to decode packets in indoor environments. Although the LoRa modulation allows the gateway to recover packets residing below the noise floor, the gateway may not be able to decode a packet residing within the interference created by severe multi-path and shadowing effects, other LoRaWAN nodes and/or networks operating in the same frequency band.
The FFT algorithm at the gateway cannot distinguish between bits 1s and 0s in the received signal because of the data availability in the undesired frequency bins or data unavailability (due to destructive interference or severe path loss) in the desired frequency bins.
For a packet reception at the gateway, we boost the decoding by creating a constructive interference of the packet using the boosters. When a constructive interference of the packet is created,
the energy levels in the desired FFT frequency bins will supersede the undesired energy levels in the other frequency bins, thereby improving the chances of decoding the chirp "0" and chirp "1" at the gateway. To be more specific, due to its {\em capture effect} capability~\cite{bankov2017mathematical, goursaud2015dedicated, el2018decoding}, a LoRa receiver (e.g., gateway/node) locks to the signal/packet that is stronger compared to the others in the same (or nearby) frequency. In summary, we ensure that a packet has the highest signal strength and may be subject to the receiver's capture effect by creating a constructive interference of the packet.


In Figure~\ref{fig:lorainup}, we explain the steps (denoted by the numbered circles) for creating the reliability boost at the gateway in the uplink. 
For better understanding, we explain the steps involving one node, one booster, and the gateway.
As shown in this figure, the booster first listens, decodes, and stores a Tx attempt of a packet by the node. 
If the gateway does not receive the packet, the node retransmits the packet (up to seven times) to the gateway. In the case of decoding error in reception, the gateway does not send an ACK, and hence the node knows that it has to retransmit the packet.
Along with the retransmission attempts by the node, the booster also transmits the {\em same packet at the same time and frequency} (i.e., channel) to the gateway, thereby creating a constructive interference (hence a capture effect) and enhancing the packet reception at the gateway. In \name{}, multiple such boosters may transmit the same packet to create a stronger constructive interference-cum-capture effect at the LoRa gateway.

\subsubsection{Boosting Energy-Efficiency}\label{sec:bee-overvew}
The LoRaWAN network server acknowledges only the first received Tx attempt of a packet (by a node) and never retransmits the ACK for the subsequent attempts of that packet. As LoRaWAN allows up to 8 Tx attempts (i.e., 7 retransmissions) of a packet, the number of wasted Tx attempts may be up to 7 if a node misses the ACK sent for the very first attempt that the gateway received. This situation causes a huge amount of energy wastage at the nodes (as per Equation (\ref{eqn:eattempts})). As a result, the lifetime of the nodes may become significantly reduced.
\begin{figure}[!htbp]
\centering 
\includegraphics[width=0.4\textwidth]{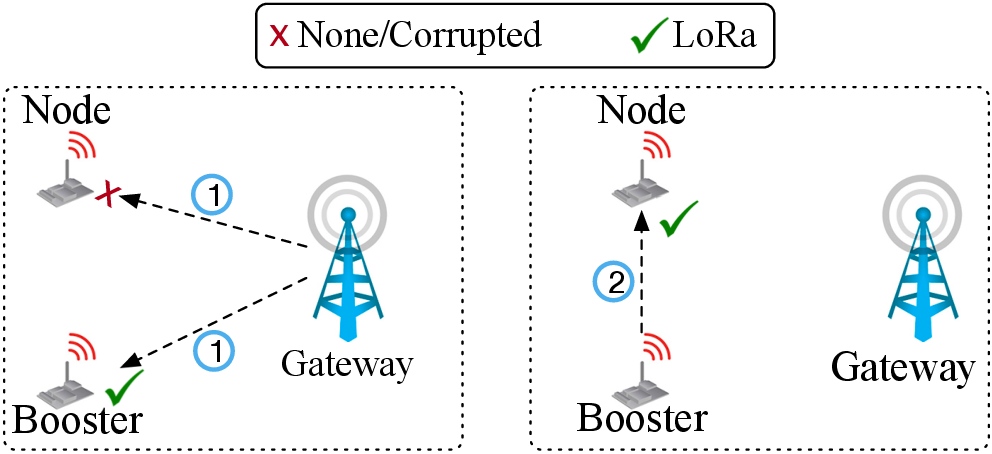} 
\caption{Boosting the ACK reception at the nodes. "x" represents no/corrupted LoRa signal/ACK, "single tick" represents a valid LoRa signal/ACK.} 
\label{fig:loraindown} 
\end{figure}
In \name{}, we enhance the energy-efficiency at the nodes by reducing the number of Tx attempts by them. Specifically, we utilize the boosters to enable ACK relay from the gateway to the nodes. Consequently, the nodes may avoid unnecessary Tx attempts as well as not waste valuable energy.

In Figure~\ref{fig:loraindown}, we explain the steps of enabling energy-efficiency in \name{} using one node, one booster, and the gateway. Specifically, we achieve energy-efficiency by relaying the missing ACKs to the nodes via the boosters. As shown in this figure, the booster first listens to the ACK (along with the node that is expecting an ACK) that is sent by the gateway. If the node misses the ACK, the booster relays the ACK to the node (step 2 in this figure). Upon reception of the relayed ACK, the node refrains from making unnecessary retransmissions of the corresponding packet, thereby reducing the energy consumption. In \name{}, multiple boosters may relay the same ACK at the same time (not shown in this figure), thereby creating a constructive interference of the relayed-ACK that may boost the capture effect at the node.

\subsubsection{Overall Workflow of the Boosters}
As discussed above, a booster participates in enhancing both the reliability at the gateway and energy-efficiency at the nodes. Therefore, it has to have a non-conflicting workflow to accommodate both of these aspects. For this, a booster maintains the following workflow. \revise{At the beginning, it hops on to different LoRaWAN uplink channels and listens for the uplink packets through ultra-low power CAD feature of LoRa (also described in our system model in Section~\ref{sec:sysmodel})}. In each uplink channel, the booster listens for a fixed duration (will be discussed in Section~\ref{sec:enable_reliability}). If it can decode any packet in any channel, it immediately starts listening for an ACK in the corresponding downlink channel for another fixed duration (will be discussed in Section~\ref{sec:ack-relay-bn}). Later, depending on the status of the packet reception at the gateway/itself or the ACK reception by the node, the booster may transmit the packet to create constructive interference at the gateway and/or relay the ACK to the node. The booster may keep repeating this workflow in between its own packets Tx to the gateway.

\subsection{Challenges in \name{}}

The boosters face several critical challenges to boost the reliability and energy-efficiency in \name{}. 

\subsubsection{Challenges in Boosting Reliability}
As shown in Figure~\ref{fig:lorainup}, a booster helps to create a constructive interference at the gateway. 
For this, it must send the same physical layer frame to the gateway along with the node at the same time and on the same channel. LoRaWAN does not provide any mechanism such that the nodes may synchronize themselves to boost each other's signals. The lack of synchronization  between a booster and a node in terms of packet, time, and channel will result in severe performance degradation at the gateway due to the additional network traffic introduced by the boosters.
Additionally, it is challenging for a booster to know if it really needs to transmit the packet (received from a node) to create a constructive interference.
It is thus very crucial that we address these challenges in \name{}.

\subsubsection{Challenges in Boosting Energy-Efficiency}
To boost the energy-efficiency at the nodes, as shown in Figure~\ref{fig:loraindown}, the boosters relay the ACKs from the gateway to the nodes. We need to address the following challenges in order to make the ACK relay beneficial for the nodes. (1) We must synchronize a booster (in terms of ACK packet, time, and channel frequency) with the ACK receive window of the desired node.
(2) We must make sure that a booster does not relay an ACK that has already been received by a node. Otherwise, this may introduce collisions with a legit ACK from another booster or the gateway for a different node in the same channel.
(3) As the boosters come into the action, it is challenging for them to know if the gateway has already sent an ACK and the corresponding node has missed it.
Otherwise, the attempts from the boosters will also be wasted and may introduce unwanted interference in the network.
In the following sections, we detail the techniques of LoRaIN. 



\subsection{Creating Constructive Interference}\label{sec:enable_reliability}
In this section, we explain our techniques for creating constructive interference at the gateway. Specifically, we explain how we synchronize the boosters and the nodes in terms of packet, time, and channel. Additionally, we discuss how a booster decides if it needs to transmit a packet for creating constructive interference or not.

\subsubsection{Packet Synchronization}\label{sec:pkt-sync}
It is crucial that a booster sends the same physical layer frame along with a node to the gateway in order to create a constructive interference. Otherwise, it may lead to an effect similar to two packets collision at the gateway, despite having a tight time and frequency synchronization between the booster and the node.
In the following, we describe how a booster receives a packet from a node, which it later transmits to the gateway. 

\begin{figure}[!htbp]
\centering  
\includegraphics[width=0.30\textwidth]{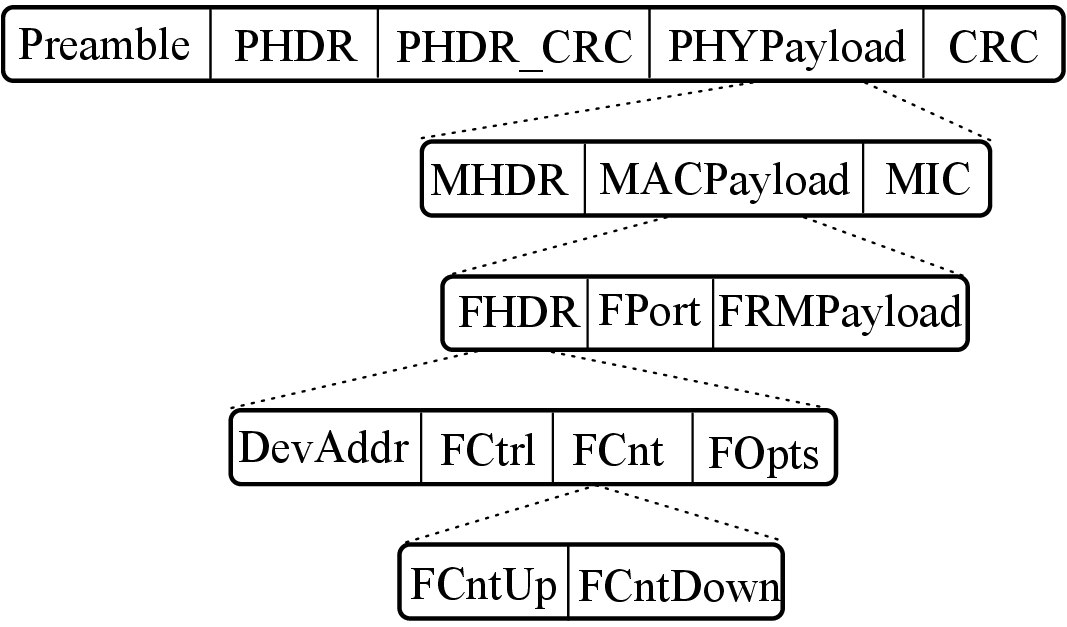}
\caption{LoRaWAN message structure (PHDR: PHY header, MHDR: MAC header, MIC: message integrity code, FHDR: frame header, DevAddr: device address, FCtrl: frame control, FCnt: frame count, FOpts: frame options)~\cite{lwan_spec103}.} 
\label{fig:packet}
\end{figure}

To receive a packet from a node, a booster first listens to the uplink Tx in the medium. The use of spread spectrum modulation in LoRa makes it impractical for the boosters to use an RSSI-based detection of signals in the medium. The reason is that the signal may reside below the noise floor. \revise{To this extent, we utilize the CAD feature of the LoRa chips, which is not used in LoRaWAN~\cite{sx1276}.} 
In CAD mode, the booster first probes for a \code{preamble} of a packet (Figure~\ref{fig:packet}) in the medium for a fixed duration. In a channel with spreading factor \code{SF} and bandwidth \code{BW}, the duration for a CAD is $(2^\text{\code{SF}} + 32) / \text{\code{BW}}$ seconds, which is approximately the duration of two LoRa symbols~\cite{sx1276}. The booster may know about the \code{SF} and \code{BW} from the gateway when requested to operate in the boosting mode. Once the booster senses an activity in the channel, it looks for the {\em start frame delimiter} (\code{SFD}) of the \code{preamble}, which is 2.25 down-chirp symbols to synchronize and receive the rest of the packet (i.e., PHY header, Header CRC, payload, and CRC). In \name{}, we use a \code{preamble} length of 10.25 symbols, which is similar to the existing LMIC LoRaWAN implementation~\cite{lmic_lora}.
The booster may have to run the CAD several times with an interval that suits its own packet Tx. Additionally, it may have to hop to different channels (as per the gateway's request) and run CAD to detect a preamble.

\subsubsection{Time Synchronization}\label{sec:uplink-time}
Once the booster has an identical copy of the packet of a node, it transmits the packet to the gateway along with the retransmission by the node. The booster, however, must synchronize in time with the node. Otherwise, the packet from the booster may create a destructive interference to the node's retransmission. Below, we describe our technique to avoid the above scenarios and create the desired constructive interference.

\begin{figure}[!htbp]
\centering 
\includegraphics[width=0.38\textwidth]{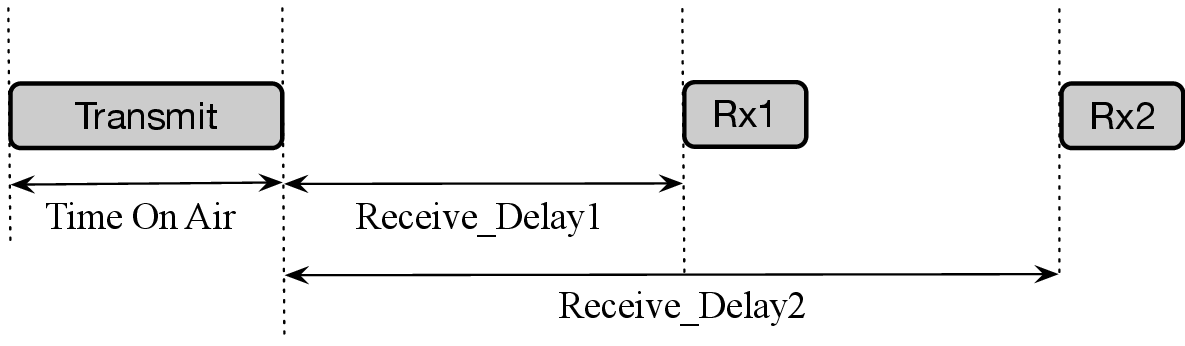} 
\caption{LoRaWAN's receive slot timing in Class-A mode of operation~\cite{lwan_spec103}.} \vspace{-0.1in}
\label{fig:rcv-slot} 
\end{figure}
To synchronize the time between a booster and a node, we utilize the node's receive slot timing window. 
As shown in Figure~\ref{fig:rcv-slot}, following an uplink Tx, the node opens two short receive windows: \code{Rx1} and \code{Rx2}. The end of \code{transmit} is the reference point for the start times of \code{Rx1} and \code{Rx2}, which are \code{Receive\_Delay1} and \code{Receive\_Delay2}, respectively. These delays are region specific and known to all the devices in the network~\cite{lwan_spec103}. Typically, $\text{\code{Receive\_Delay2}} = \text{\code{Receive\_Delay1}} + 1$ seconds. While datarate for \code{Rx1} is fixed and identical to the \code{transmit} datarate, the datarate for \code{Rx2} is region-specific and known to all the devices. If the node does not receive an ACK in any of these windows, it retransmits the packet after $\text{\code{Receive\_Delay2}} + \tau$ seconds, where $\tau$ is the required duration (region-specific) to detect and receive an ACK in \code{Rx2}. A booster also follows the same timing and transmits the packet.

\subsubsection{Channel Synchronization}\label{sec:ch-sync}
The boosters must transmit in the same channel (i.e., frequency) as a node to create the constructive interference. Otherwise, the packets from the boosters will interfere the ongoing Tx in the undesired channels. To synchronize the channel between a booster and a node, we follow the default channel increment procedure of LoRaWAN. In LoRaWAN, a node retransmits a packet on $Channel_{\text{curr}} = (Channel_{\text{prev}} + 1) \bmod N$, where $Channel_{\text{prev}}$ is the channel in which the last Tx was lost and $N$ is the number of uplink channels the LoRaWAN gateway is capable of listening to. Since the booster knows $Channel_{\text{prev}}$ during the packet synchronization, it calculates the desired channel based on the above equation and transmits to the gateway in order to create a constructive interference.

\subsubsection{Decision on the Reliability Boost}
It is critical for a booster to know if it needs to transmit a packet to create the constructive interference, despite being able to synchronize with the packet, time, and channel of a node. We achieve this in the boosters using the following technique. A node encodes 3 bits of additional information (to represent 8 Tx attempts) in the last octet (out of 15) of the \code{FOpts} field (Figure~\ref{fig:packet}), which is unused in LoRaWAN~\cite{lwan_spec103}.
Specifically, we use the flow of 3-bit natural binary numbers to represent the Tx attempt count. For example, we use 000 to denote the first Tx attempt, 001 to denote the second Tx attempt, and so on. 
Upon receiving a packet, a booster checks this information. Later, if it finds that no ACK is sent in \code{Rx1} or \code{Rx2} (Figure~\ref{fig:rcv-slot}) for the node and the node has not exhausted all the attempts yet, it transmits the packet at the specified time and channel.

\begin{figure*}[!htbp] 
    \centering 
      \subfigure[Only two transmitters are active\label{fig:ci_tx}]{
        \includegraphics[width=.28\textwidth]{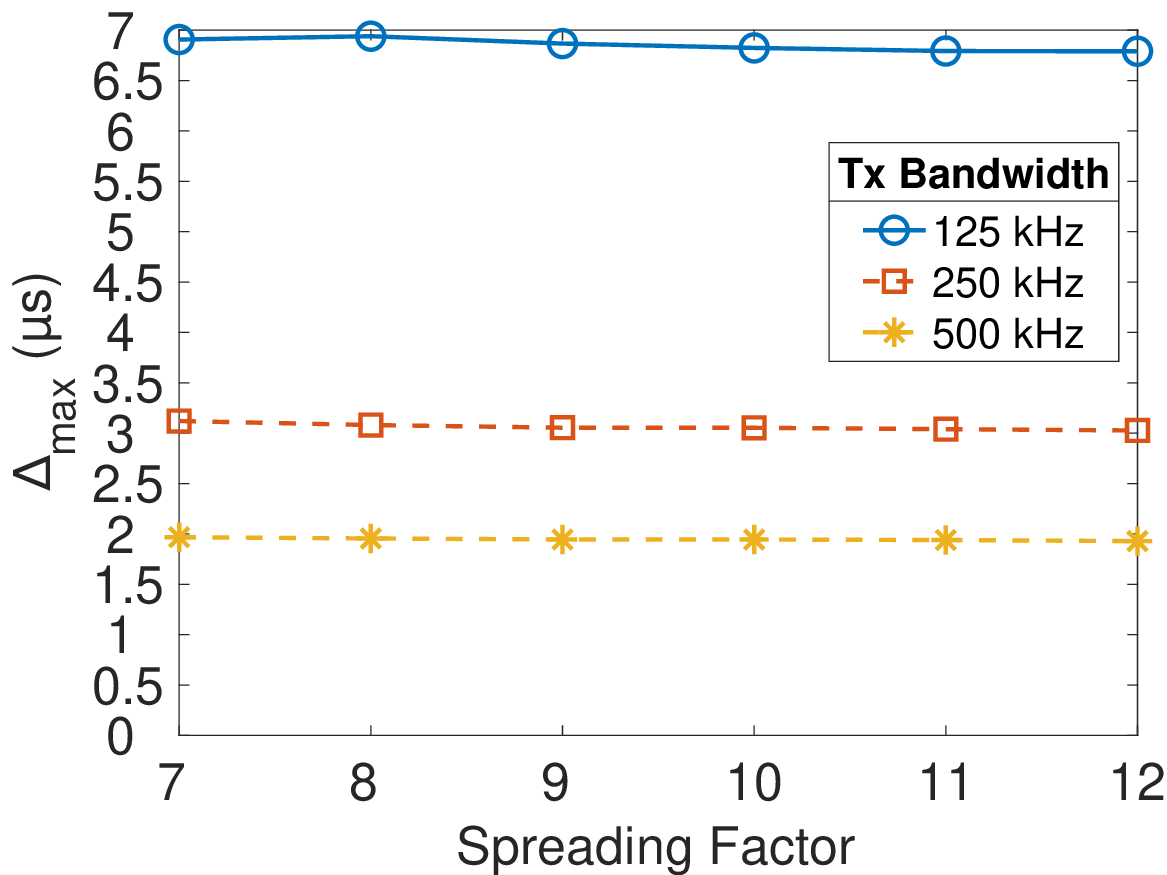}
      }\hfill
      \subfigure[Spectrogram and superimposition (LoRa chip is oversampled 20x for better visibility) \label{fig:ci_specto}]{
    	\includegraphics[width=0.28\textwidth]{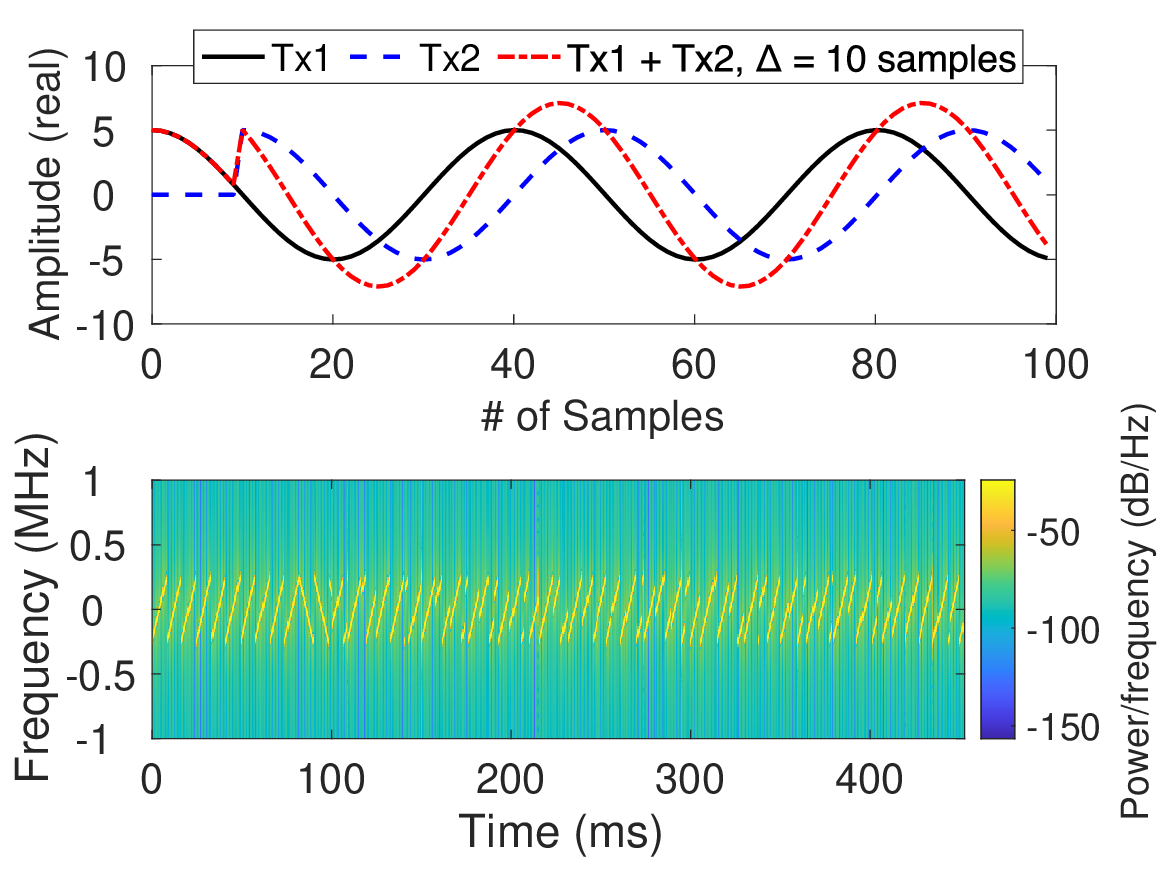}
      } \hfill
      \subfigure[Two transmitters and an interferer are active\label{fig:ci_inf}]{
        \includegraphics[width=.28\textwidth]{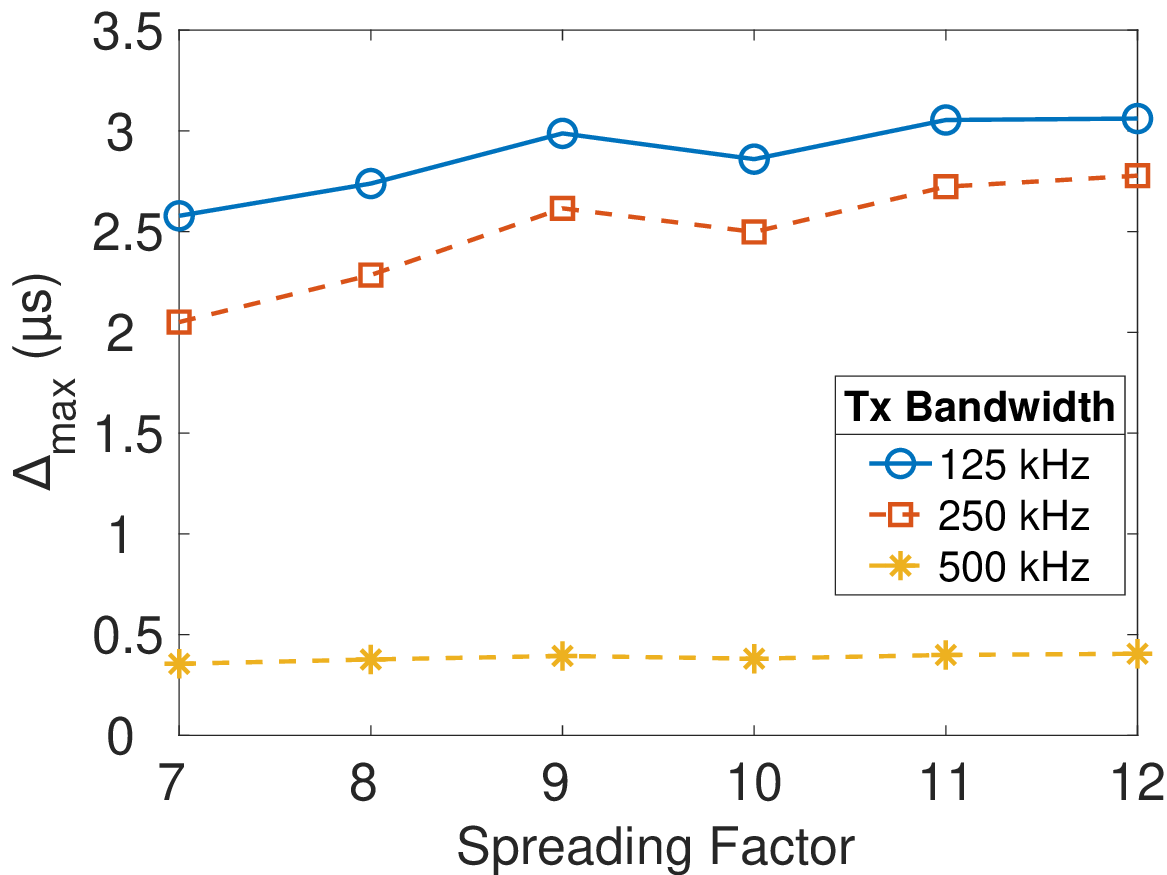}
      }
    \caption{Constructive interference timing requirement analysis.} 
    \label{fig:ci_sim}
\end{figure*}
\subsection{Enabling Acknowledgment Relay}\label{sec:ack-relay-bn}
In the following, we detail our techniques to overcome the challenges related to the ACK relay in \name{}. 
Specifically, we discuss how the boosters synchronize the time and channel to relay the ACKs as well as avoid relaying duplicate ACK or those that may introduce false negatives in the network.

\subsubsection{ACK Packet, Time, and Channel Synchronization}
It is crucial for a booster to correctly receive an ACK and relay to a node at the exact time and channel as expected. Otherwise, the relayed ACK may not be decoded correctly at the nodes and/or collide with other ACKs in the network, which will increase the number of Tx attempts by the nodes. For this, we again utilize the LoRaWAN node's receive slot timing (Figure~\ref{fig:rcv-slot}) to receive the ACK packet and synchronize the time of relay between a booster and a node.
Since the booster is already synchronized with the \code{transmit} window (Section~\ref{sec:pkt-sync}), it can receive the ACK in \code{Rx1} or \code{Rx2} from the gateway and relay it during the next \code{Rx1} and \code{Rx2} slots of the node. To synchronize the channel, we use the default channel mapping of the Tx-Rx operations in LoRaWAN, which is defined for \code{Rx1} as $Channel_{\text{Rx}} = Channel_{\text{Tx}} \bmod 8$, where 8 is the number of downlink channels in LoRaWAN. For \code{Rx2} Tx-Rx channel mapping, we use the first downlink channel that is also used in the existing LMIC LoRaWAN implementation~\cite{lmic_lora}. Also, We use the respective datarates of \code{Rx1} and \code{Rx2}, as discussed in Section~\ref{sec:uplink-time}. Upon receiving an ACK, a node checks for the \code{DevAddr} and \code{MIC} fields (Figure~\ref{fig:packet}) to check if the ACK is intended for it or not. {If multiple boosters relay the same ACK to a node, it will create a constructive interference, and hence the node has even higher chances of receiving the ACK due to its capture-effect capability.}


\subsubsection{Handling Duplicate ACKs}
While relaying an ACK, a booster should avoid sending duplicate copies of the same ACK that has already been received by a node. Otherwise, that may collide with and destroy a legit ACK for another node on the same channel.
Such duplication may be introduced because the node has already received the ACK from the gateway or a booster. To avoid this, a booster compares two packets received in the latest two consecutive \code{Transmit} windows (Figure~\ref{fig:rcv-slot}). Specifically, the booster compares both the \code{FCntUp} and 3 bits of ( that are encoded by the node) \code{FOpts} fields. A node increases \code{FCntUp} by one only for each packet with new payload data but keeps it the same for a retransmission attempt of an older packet. Thus, a booster relays an ACK only if the latest two \code{FCntUp} fields (of those packets) are the same and the Tx count (on \code{FOpts}) of the latest packet is greater than that in the second latest packet.


\subsubsection{Handling Missing ACKs}
There may be a few cases where the boosters may not receive any ACK from the gateway to relay to a node. This may happen when the gateway sends the ACK and the node misses it 
before any of the boosters come into action.
In this case, we allow the node to increment \code{FCntUp} by one and retransmit the same packet. As per the LoRaWAN specification, the gateway sends a new ACK for an increase in the \code{FCntUp} field of the packet. The trade-offs of this technique include at most 7 wasted Tx attempts by the boosters of the packet with the old \code{FCntUp}.


\subsection{\name{} Timing Requirements}\label{sec:timing-analysis}
\revise{In this section, we analyze the timing requirements for the constructive interference in \name{} both empirically through Matlab simulations and mathematically.
Overall, both the analyses are consistent with each other.}

\subsubsection{Empirical Analysis}\label{sec:empirical_analysis}
In this section, we present our empirical analysis of the timing requirements for the constructive interference in \name{}. Specifically, we perform Matlab simulations to evaluate the maximum allowable temporal displacement (say, $\Delta_{max}$) between two LoRa transmitters (e.g., a node and a booster transmitting the same payload at the same frequency) such that they interfere constructively. In \name{}, even if a node and a booster transmit at the same time, their signals may reach the gateway with a temporal offset $\Delta$ due the difference in their distances from the gateway. 
We use the Matlab LoRa simulator developed by the authors in~\cite{loramatlab}. In simulations, we analyze $\Delta_{max}$ for different \code{SF}s and \code{BW}s while the \code{CR} is fixed at $\frac{4}{5}$ for a payload of 30bytes from each transmitter with a Tx power of 14dBm. We also add white Gaussian noise to the superimposed signal. The signals from both transmitters have the same amplitude, but one of them is delayed by a variable $\Delta$ with 10ns granularity in the interval [0, $\frac{1}{\code{BW}}$], where $\frac{1}{\code{BW}}$ is the chip duration in a LoRa symbol (e.g., for 125kHz \code{BW}, 125000 chips/s or 8$\mu$s/chip while the symbol is 2$^{\code{SF}}$ chips long). In simulations, we analyze the behavior of $\Delta_{max}$ for two cases: (1) only two transmitters are active and (2) an interferer is active as well with the two transmitters. For each pair of \code{BW} and \code{SF} in these cases, we run 100 simulations with different seeds for the noise.

Figure~\ref{fig:ci_tx} shows the $\Delta_{max}$ (averaged over 100 runs) for successful constructive interference (100\% PRR) for \code{BW}s 125, 250, and 500kHz with \code{SF}s 7--12 when only two transmitters are active in the same frequency. As shown in this figure, the $\Delta_{max}$ for a given \code{BW} is almost constant regardless of the change in the \code{SF}. For 125kHz \code{BW}, the average $\Delta_{max}$ over \code{SF}s 7--12 is 6.85$\mu$s with a standard deviation of 0.061$\mu$s. For 250 and 500kHz \code{BW}s, the average $\Delta_{max}$'s are 3.06 and 1.95$\mu$s, respectively, with standard deviations 0.033 and 0.013$\mu$s, respectively. Our observation is that the $\Delta_{max}$ for any pairs of \code{BW} and \code{SF} stay {\em considerably} below the corresponding chip duration. For 125, 250 and 500kHz \code{BW}s, the chip durations are 8, 4, and 2$\mu$s.

\revise{Figure~\ref{fig:ci_specto} shows the spectrogram (bottom) and time-domain signals (top) of two active transmitters with a \code{BW} of 500kHz, \code{SF} of 10, and $\Delta$ of 10 samples ($\approx 1\mu$s) with oversampling a chip 20x to provide better visibility in representation, where we have a successful constructive interference.} For this configuration, $\Delta_{max}$ = 1.9441$\mu$s. For a \code{BW} of 125kHz and \code{SF} of 10 (most common configuration in Section~\ref{sec:rationale} chosen by ADR), $\Delta_{max}$ = 6.8243$\mu$s. Figure~\ref{fig:ci_inf}, on the other hand, shows that the $\Delta_{max}$ exhibits a bit of randomness for different pairs of \code{BW} and \code{SF} when an interferer is active (i.e., transmitting a different payload) along with the two transmitters. For 125, 250, and 500kHz \code{BW}s, the average $\Delta_{max}$'s over \code{SF}s 7--12 are 2.88, 2.49, and 0.385$\mu$s, respectively, with standard deviations 0.193, 0.278, and 0.017$\mu$s, respectively. This figure also shows that $\Delta_{max}$ stays considerably below half-chip duration with the exception for \code{BW} = 250kHz in the case of two transmitters and an interferer. Overall, our simulations show that we may observe a successful constructive interference of two LoRa transmitters when $\Delta_{max} < \frac{1}{\code{BW}}$.

\subsubsection{Theoretical Analysis}
\revise{In this section, we present an intuitive mathematical analysis of the timing requirements in \name{} for the desired packets to collide constructively. For the sake of simplicity, we consider only two transmitters, a \name{} node and a booster, in our analysis. Additionally, we focus on the constructive inference of two identical LoRa symbols, one transmitted from the node and the other from the booster, in accordance with the \name{} design principles and symbol-level mathematical expressions presented in Section~\ref{sec:lora-phy}.
However, this analysis may be easily extended to two identical LoRa frames (i.e., packets) as a LoRa frame is a collection of LoRa symbols transmitted consecutively one after another, e.g., for $L$ symbols, the transmitted frame $x[n] = \sum_{l=0}^{L-1}s_{\alpha_l}[n \mod M]$.}

\revise{Now, let us consider that the transmitted symbol $\alpha$ from the booster is delayed by $n_i$ samples or LoRa chips at the \name{} receiver compared to the symbol $\alpha$ transmitted by the node at sample instance $n$. Consequently, these two symbols may be represented as $s_{\alpha}[n - n_i]$ and $s_{\alpha}[n]$, respectively. Their mathematical expressions can be derived easily from Equation (\ref{eqn:lora_symbol}). For the sake of this intuitive proof, we may omit their mathematical expressions here. In terms of the corresponding temporal displacement between these two symbols, we may have $\Delta = (n - n_i)T_s$, where $T_s$ is the sampling period in discrete-time LoRa symbol waveforms and $\Delta$ has been used as the temporal displacement in our empirical analysis in Section~\ref{sec:empirical_analysis}. For the constructive inference to occur at the receiver while it is locked to the receive path of the symbol from the node (or booster), the symbol from the booster (or node) may not be delayed by more than the duration of a single chip (or a sample in this case). If the temporal displacement is more than the duration of one chip, then the capture effect may not be in effect in the \name{} receiver, leading to the scenario similar to the adverse effects of inter-symbol interference (from two different transmitters in this case). Consequently, the maximum temporal displacement $\Delta_{max}$ between the two superimposed symbols may not exceed the duration discussed above. With a symbol period $T$ and $M = 2^{\code{SF}}$ samples (i.e., chips) per symbol period (Section~\ref{sec:lora-phy}), we have $\Delta_{max} < \frac{T}{2^{\code{SF}}}$. We know that in LoRa CSS modulation, $T = \frac{2^{\code{SF}}}{\code{BW}}$, and thus $\Delta_{max} < \frac{1}{\code{BW}} = T_s$.}

\subsection{Discussion on Security}\label{sec:security}
The security aspect of \name{} is out of the scope of this paper. We, however, provide a brief discussion below on the security of LoRaIN for the implementation used in this paper. To enable security in peer-to-peer packet/ACK receptions in LoRaIN, e.g., between nodes and boosters, or other communications (e.g., gateway to booster ACK reception on behalf of the other nodes), we use the same security keys (e.g., network key, application key, and application identifier) across all the devices (e.g., gateway, nodes, boosters) in the network. Additionally, the boosters learn about the \code{DevAddr} fields of the nodes through the gateway, which is done during the bootstrapping of the network or when the gateway asks a node to operate in the boosting mode. We leave the study on randomizing these keys and securing these key exchanges as a future work.

\begin{figure*}[t] 
    \centering 
      \subfigure[CAD detection accuracy\label{fig:cad_det}]{ 
    \includegraphics[width=0.28\textwidth]{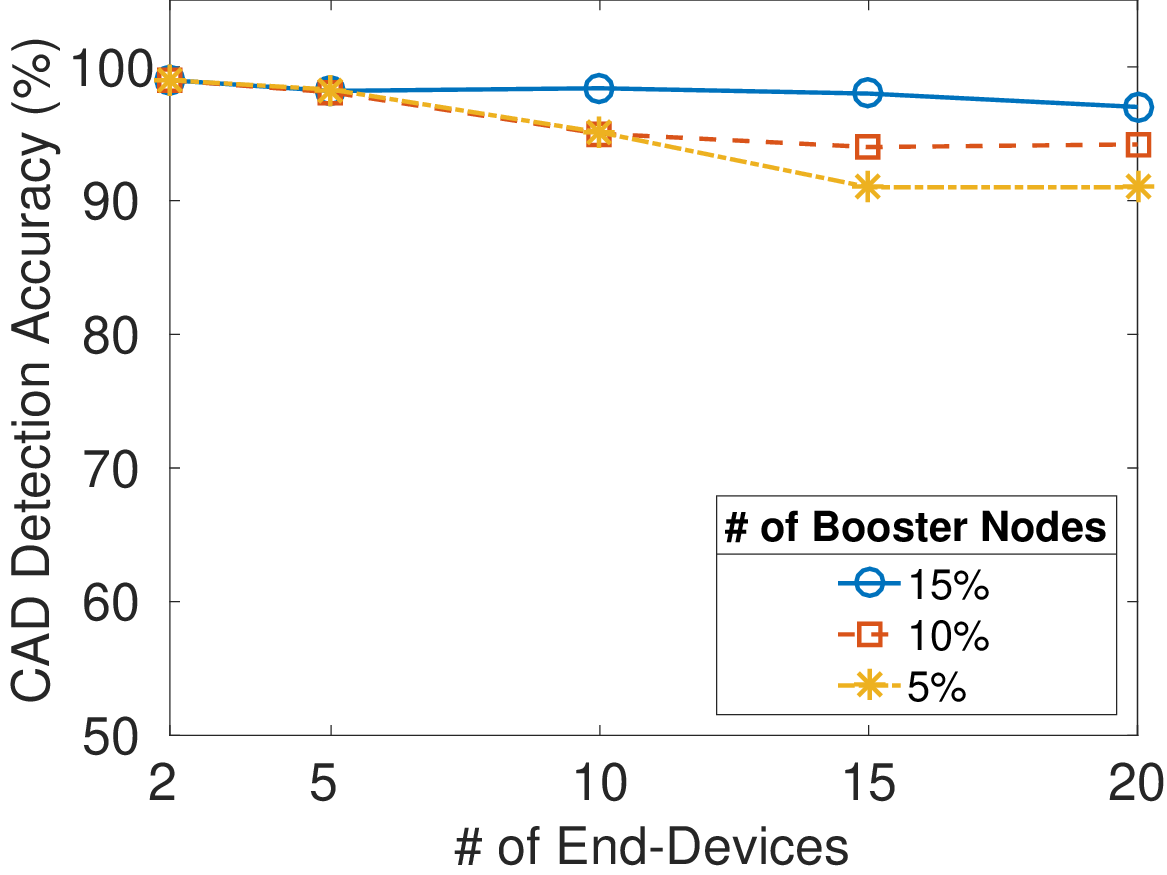}
      }\hfill 
      \subfigure[CAD reception accuracy\label{fig:cad_rec}]{
        \includegraphics[width=.28\textwidth]{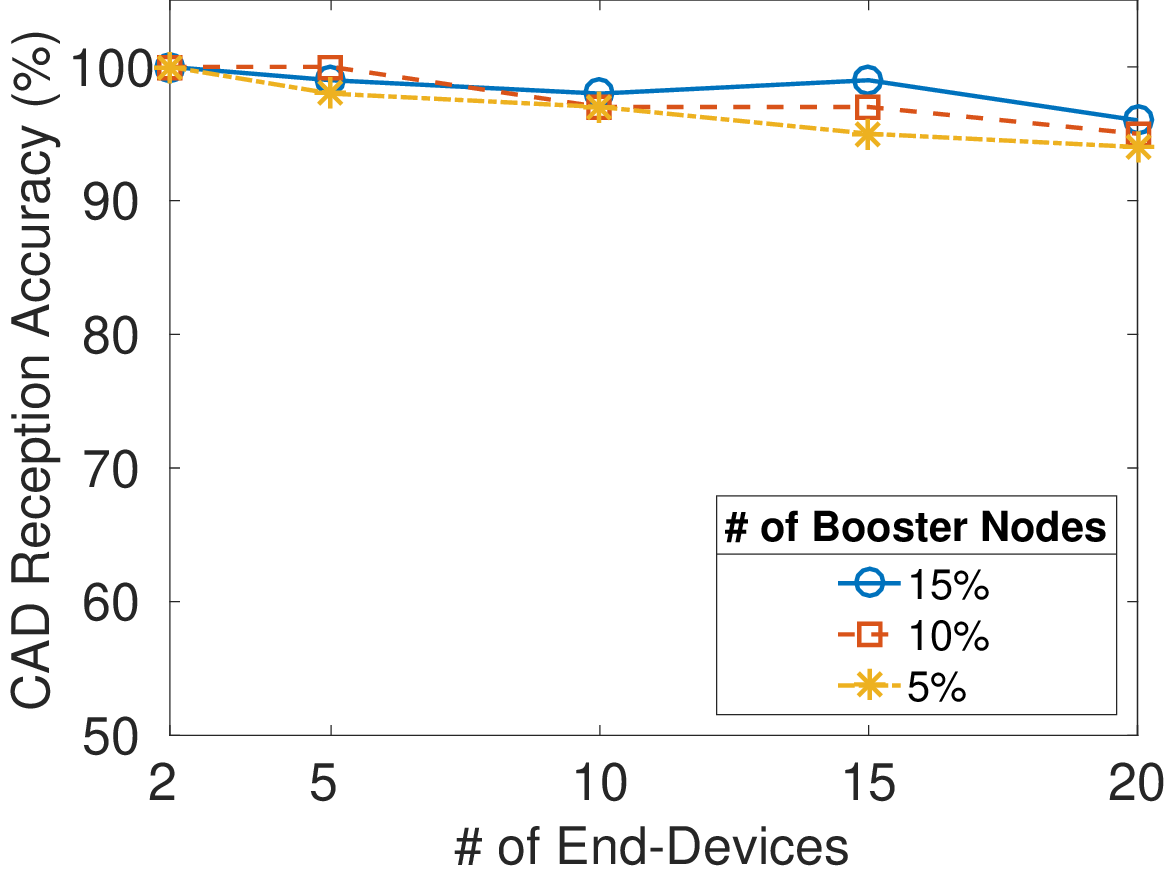}
      }\hfill 
      \subfigure[Energy overhead at boosters\label{fig:booster_energy}]{
        \includegraphics[width=.28\textwidth]{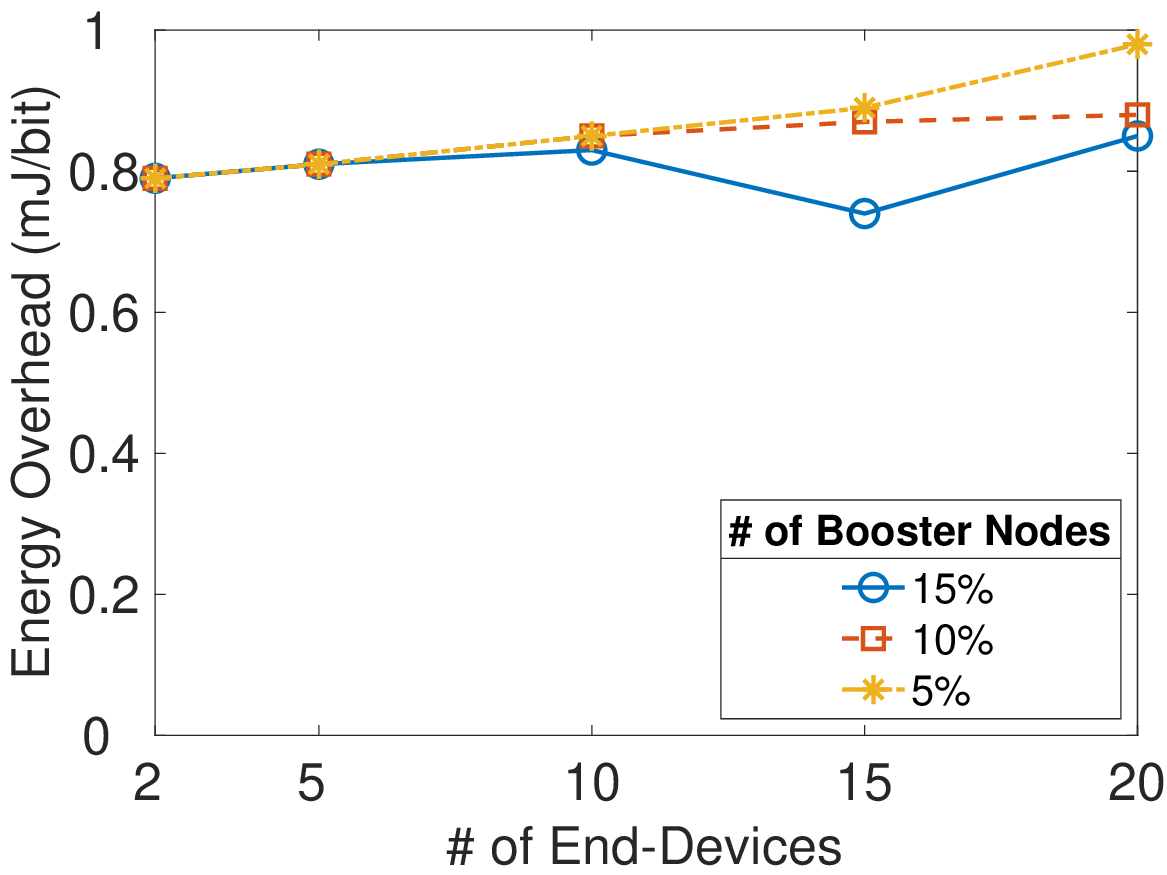}
      } 
    \caption{Performance of \name{} CAD.} 
    \label{fig:cad_exp}
\end{figure*}
\section{Experiment}\label{sec:experiment}
In this section, we evaluate the performance of \name{} through extensive experiments in the same indoor area of approximately 600ft$^2$, as depicted in Figure~\ref{fig:indoor}. 
In the following, we first discuss our experimental setup and then present the performance of \name{} CAD (including energy overhead at the boosters), protocols for the reliability and energy boost, and network performance.

\subsection{Implementation and Default Setup}\label{sec:implementation}
We implement \name{} using one LoRaWAN gateway and 20 LoRaWAN end-devices (i.e., nodes). The gateway is a RAK2245 Pi Hat that runs on a Raspberry Pi and can simultaneously receive on 8 uplink channels~\cite{rak_gateway}. We use the ChirpStack LoRaWAN network server that controls the gateway and runs locally on the gateway~\cite{cstack}. We use the Dragino LoRa Hat (SX1276 LoRa chip) on Raspberry Pi as the nodes~\cite{dragino, sx1276}.
We also customize the LMIC 1.6 LoRaWAN development library and configure it with the parameters (e.g., frequency in 915MHz band, 125KHz uplink channels, 500kHz downlink channels, and maximum usable SF of 10) that are specific to our region~\cite{lmic_lora, lwan_spec103} and match the capability of our gateway (RAK2245 on a Raspberry Pi). In experiments, 
we choose between 5\% and 15\% of the nodes to act as the boosters while we vary the total number of nodes between 2 and 20.
We calculate the actual number of boosters as $\ceil{P_{BN} \times M}$, where $P_{BN}$ is the percentage of nodes acting as boosters and $M$ is the total nodes.
For example, if 15 nodes are active in the network and we want 15\% of them to act as boosters, then the actual number of boosters is $\ceil{0.15 \times 15} = 3$. 
We use this technique so that we may evaluate \name{} under varying number of nodes and depict easily. 
In each of the experiments, a node (including each booster) sends 100 confirmed packets with an inter-packet interval of 1 minute. Each packet contains a random payload of 30 bytes. 
Unless stated otherwise, these are our default parameter settings for all the experiments presented henceforth.

\subsection{\name{} Carrier Activity Detection}
In this section, we evaluate the performance of the activities related to \name{} CAD. 
Specifically, we look into the CAD detection accuracy, reception accuracy, and energy overhead at the boosters. {\em CAD detection accuracy} is defined as the ratio of the number of packets (including ACKs) whose \code{preambles} are detected and synchronized by the boosters to the total number of packets sent by the nodes or gateway. {\em CAD reception accuracy} is defined as the ratio of the number of correctly received packets (including ACKs) at the boosters to the total number of packets sent by the nodes or gateway. To calculate the {\em energy overhead} (in joule per bit unit) at the boosters, we take into account the total energy spent by the boosters in CAD detection, CAD reception, and one-shot forwarding of the packets (including ACKs) to the nodes or the gateway.

\subsubsection{CAD Detection Accuracy}
Figure~\ref{fig:cad_det} shows the CAD detection accuracy of \name{} while the number of nodes and boosters are varied between 2 and 20 and between 5\% and 15\%, respectively. As shown in this figure, for 2 nodes and 5\% boosters, the CAD detection accuracy is as high as 99\%. Also, it increases with the increase in the number of boosters. For example, in the case of 20 nodes with 5\%, 10\%, and 15\% boosters, the CAD detection accuracies are approximately 91\%, 94.2\%, and 97\%, respectively. This experiment thus shows that the CAD detection accuracy is very high in \name{}, which confirms that boosters nodes are capable of detecting and synchronizing with almost all the packets in the network.

\subsubsection{CAD Reception Accuracy}
Figure~\ref{fig:cad_rec} shows the CAD reception accuracy as we vary the number of nodes and boosters. For 2 nodes and 5\% boosters, the CAD reception accuracy is 100\%. Overall, as we increase the number of boosters, the CAD reception accuracy also increases. For example, in the case of 20 nodes with 5\%, 10\% and 15\% boosters, the CAD reception accuracies are approximately 94\%, 95\%, and 97\%, respectively. Such high CAD reception accuracy is very crucial in \name{} since constructive interferences are created by the boosters using these packets.

\subsubsection{Energy Overhead at Boosters}\label{sec:booster_energy}
Figure~\ref{fig:booster_energy} depicts the energy consumption of the boosters in mJoule/bit unit when the number of nodes (between 2 and 20) and boosters (between 5\% and 15\%) is varied, which may be considered overhead if they are battery-powered. As shown in this figure, if we increase the number of boosters (for the same number of nodes), the average energy overhead tends to decrease gradually. For 5 nodes with 5\%, 10\% and 15\% boosters, the average energy overhead is 0.81, 0.81, and 0.81 mJ/bit, respectively. For 10 nodes with 5\%, 10\%, and 15\% boosters, it is 0.83, 0.85, and 0.85 mJ/bit, respectively. For 20 nodes with 5\%, 10\%, and 15\% boosters, it is 0.86, 0.88, and 0.98 mJ/bit, respectively. On the other hand, the average energy overhead increases gradually if we increase the number of nodes while the number of boosters is fixed (as shown in this figure). While we focus only on booster's energy overhead in this section, we detail the overall network (including booster nodes) energy consumption later in Section~\ref{sec:network-latency-energy} that shows that \name{} with energy overhead at the boosters may still consume 2.5x less than the overall energy consumption in LoRaWAN.


\begin{figure}[!htbp]
\centering 
\includegraphics[width=0.28\textwidth]{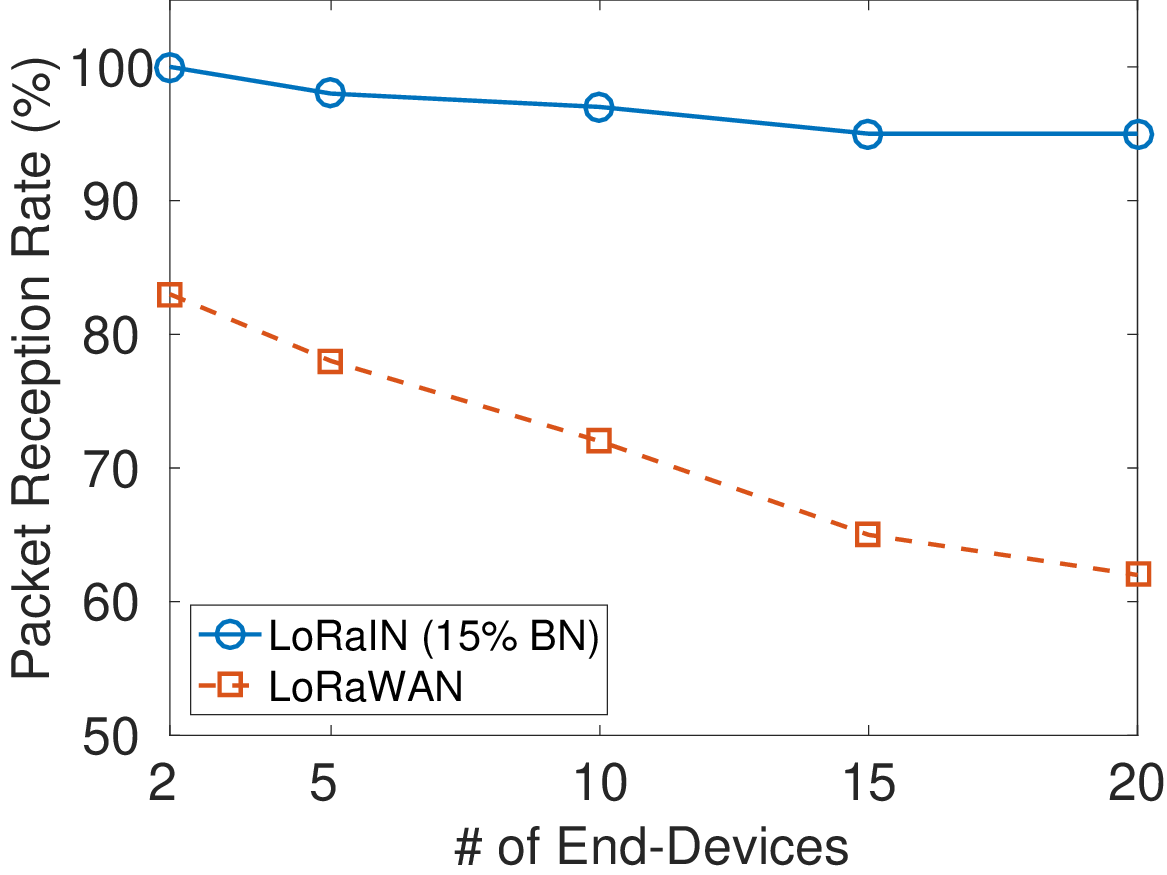} 
\caption{Reliability at the gateway.} 
\label{fig:prr_gw}
\end{figure}
\subsection{Experiment on Constructive Interference}
In this section, we evaluate the performance of the \name{} boosters in terms of creating successful constructive interferences at the gateway. Specifically, we calculate the PRR (packet reception rate) at the gateway. To recall, packet reception rate is the ratio of the number of packets received at the gateway to the total number of packets sent by the nodes. 

\subsubsection{Results}
Figure~\ref{fig:prr_gw} shows the PRR at the gateway as we vary the number of nodes from 2 to 20 and keep the number of boosters fixed at 15\%. In this figure, we compare the performance of \name{} with LoRaWAN as well. In the case of 2 nodes, the PRR at the gateway is 100\% in \name{}, compared to 83\% in LoRaWAN. As we increase the number of nodes, the performance difference between \name{} and LoRaWAN becomes more prominent. Particularly, as the number of nodes increases, the PRR at the gateway in LoRaWAN goes down sharply, while it is still very high in \name{}. For example, in the case of 20 nodes, the PRR at the gateway is 95\%, compared to only 62\% in LoRaWAN. This experiments thus demonstrate that \name{} is much more reliable in indoor, compared to LoRaWAN.

\subsection{Experiments on Acknowledgment Relay}
In this section, we evaluate the performance of \name{} in terms of ACK relays from the boosters. We calculate the PDR and average number of Tx attempts per packet at the nodes. PDR is the ratio of the number of acknowledged packets to the number of total packets sent. If the ACK relays by the boosters work, the PDR should increase and the average number of Tx attempts should decrease at the nodes.
\begin{figure}[!htbp] 
    \centering 
      \subfigure[PDR at the nodes\label{fig:pdr_ed}]{ 
        \includegraphics[width=.28\textwidth]{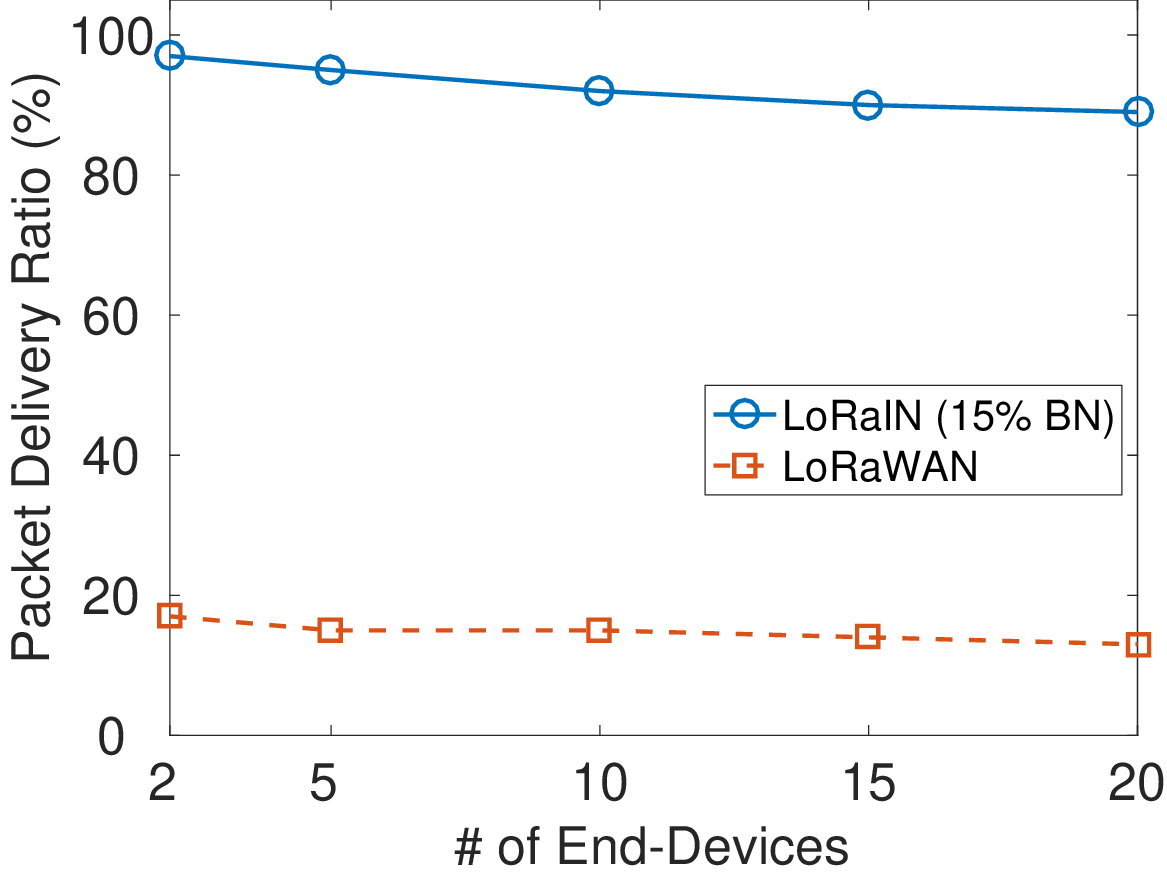}
      }\vfill 
      \subfigure[\# of Tx attempts by the nodes\label{fig:tx_attempt_ed}]{
        \includegraphics[width=.28\textwidth]{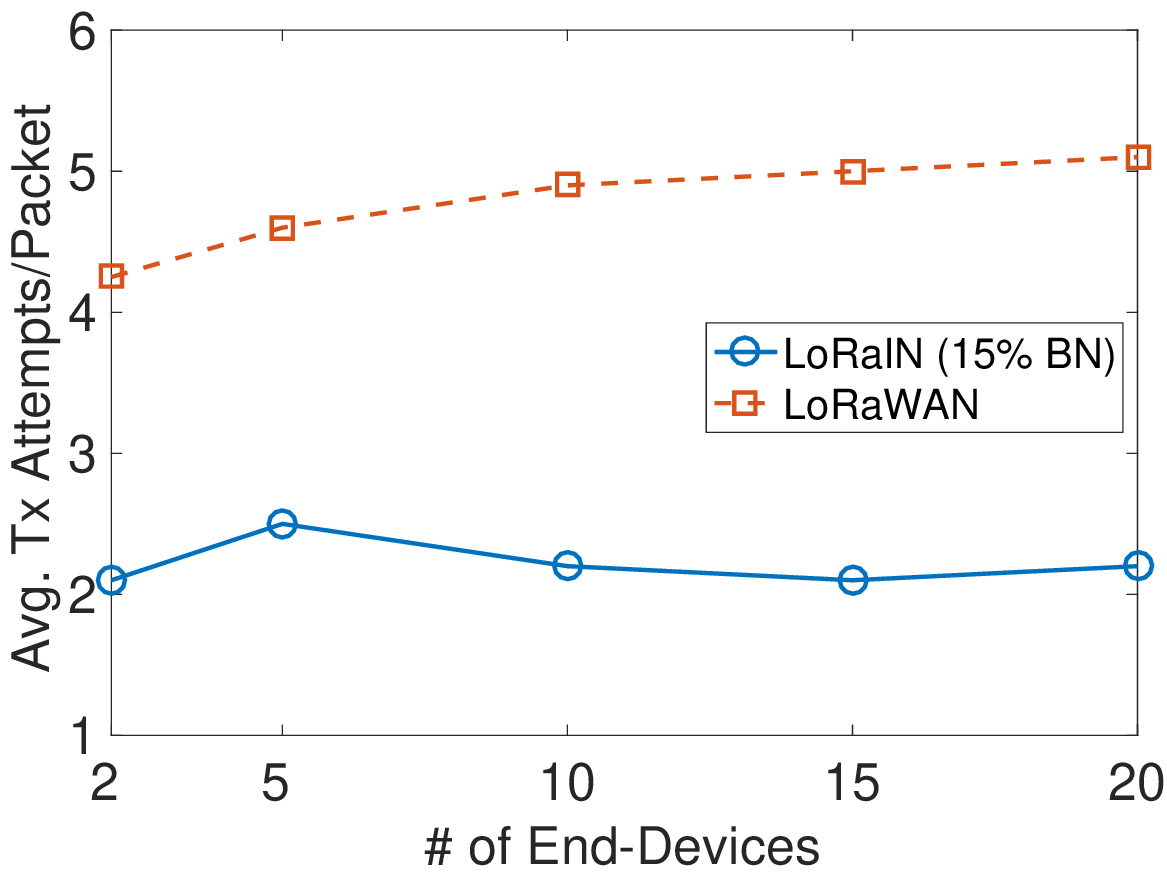} 
      } 
    \caption{Performance in terms of ACK relay.} 
    \label{fig:lorain_exp}
\end{figure}

\begin{figure*}[!htbp] 
    \centering 
      \subfigure[Effective bitrate at the gateway\label{fig:bitrate_gw}]{
    \includegraphics[width=0.28\textwidth]{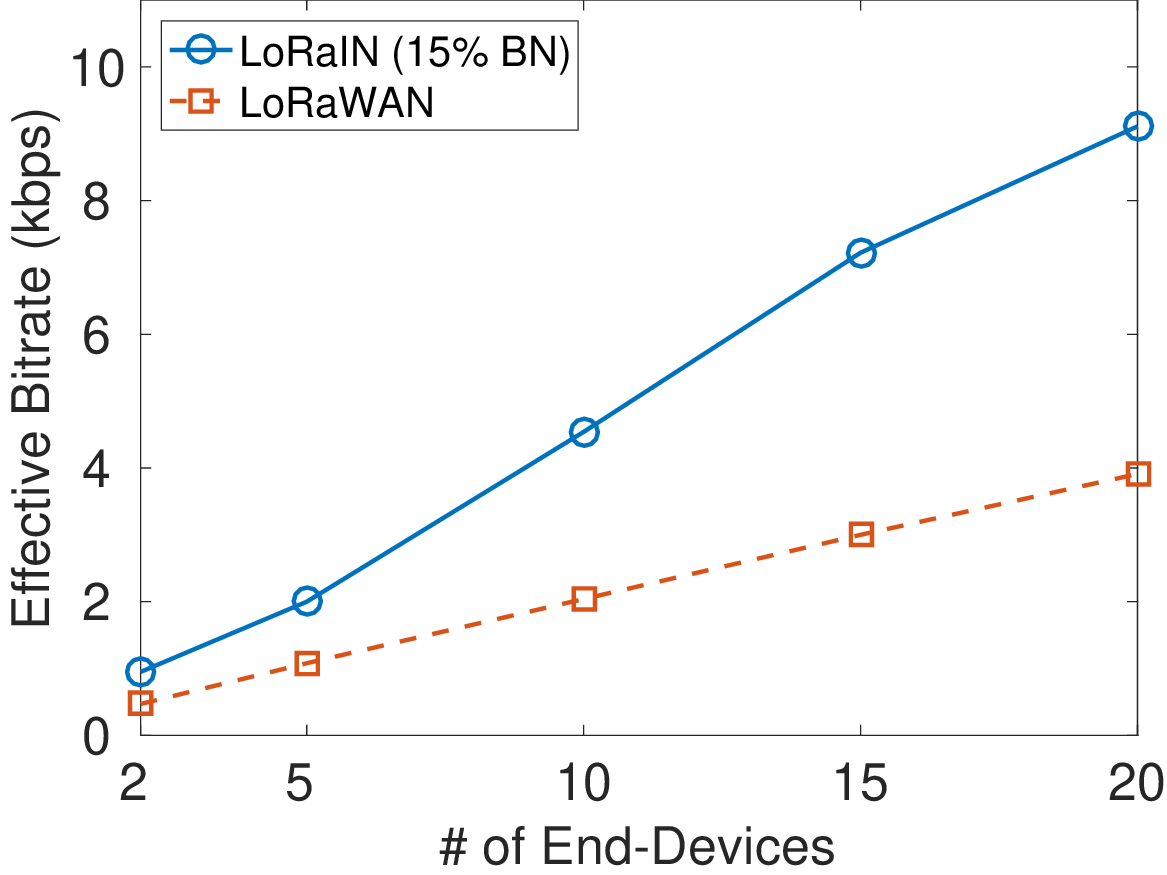}
      }\hfill
      \subfigure[Avg. latency/packet at the nodes\label{fig:latency_ed}]{
        \includegraphics[width=.28\textwidth]{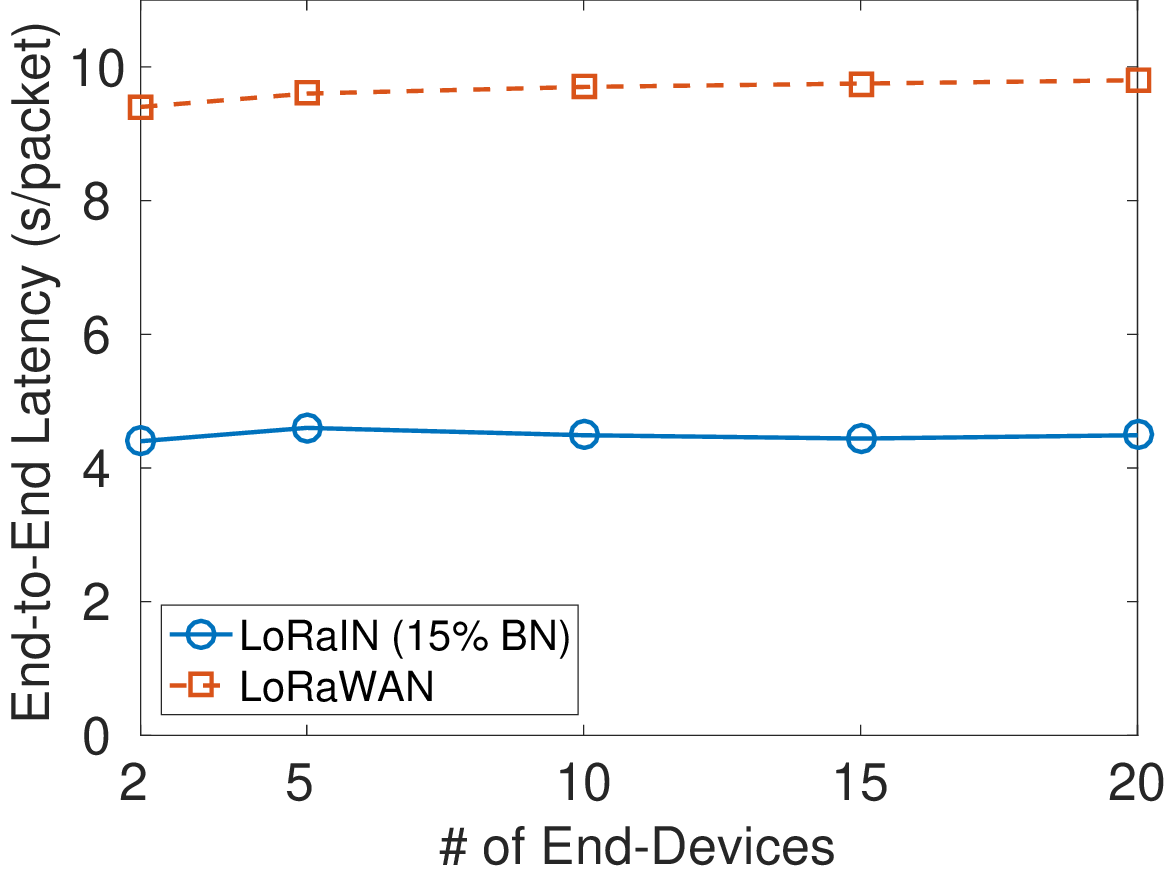}
      }\hfill
      \subfigure[Avg. energy consumption/packet at the nodes (including boosters)\label{fig:energy_ed}]{
        \includegraphics[width=.28\textwidth]{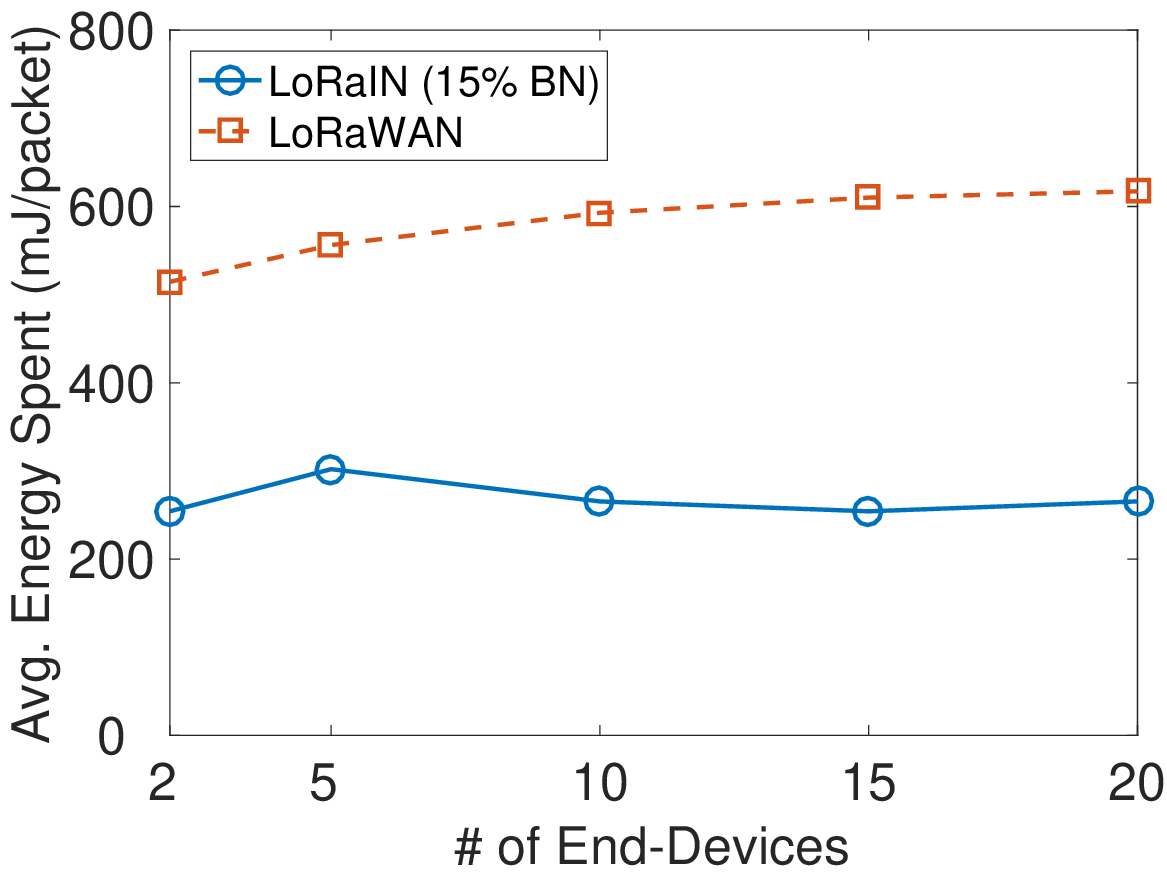}
      } 
    \caption{Indoor network (including boosters) performance evaluation.} 
    \label{fig:net_exp}
\end{figure*}
\subsubsection{Packet Delivery Ratio}
Figure~\ref{fig:pdr_ed} shows the PDR at the nodes as we vary the number of nodes between 2 and 20 and keep the boosters fixed at 15\%. Additionally, we compare this performance with LoRaWAN. As shown in this figure, for all the cases, the PDR in \name{} is very high, compared to those in LoRaWAN. For example, in the case of 20 nodes, the PDR in  \name{} is approximately 90\%, compared to only 13\% in LoRaWAN. This experiment thus demonstrates that the ACK relays by the boosters are very effective in \name{}, which outperforms LoRaWAN significantly.

\subsubsection{Transmission Attempts per Packet}
Figure~\ref{fig:tx_attempt_ed} shows the average number of Tx attempts per packet by the nodes in \name{} and also compares with LoRaWAN. In this experiment, the average number of Tx attempts per packet by the nodes almost stays the same in \name{}, compared to a noticeable increase in LoRaWAN, as we increase the number of nodes from 2 to 20 while keeping the number of boosters fixed at 15\%. For example, in the case of 2 and 20 nodes, the average numbers of Tx attempts per packet in \name{} are 2.1 and 2.2, compared to 4.2 and 5.2 in LoRaWAN, respectively. As shown in this experiment, \name{} performs more than twice as better as LoRaWAN, thereby showing the feasibility of its ACK relay.

\subsection{Network Performance Analysis}
In this section, we evaluate the network performance in terms of effective bitrate, average end-to-end (E2E) latency, and energy consumption. {\em Effective bitrate} is calculated based on the distinct packets received at the gateway. {\em End-to-end latency} per packet is defined as the time difference between the start of the first Tx attempt of the packet and end of its ACK reception. We calculate the {\em average energy consumption} per packet at the nodes considering the E2E latencies (energy/packet $\propto$ E2E/packet) of their packets.

\subsubsection{Effective Bitrate}
Figure~\ref{fig:bitrate_gw} shows the effective bitrate at the gateway for all the packets sent from the nodes. In this experiment, we vary the number of nodes between 2 and 20 and compare the performances between \name{} and LoRaWAN. As shown in this figure, the effective bitrate at the gateway increases at a higher speed in \name{}, compared to the bitrate in LoRaWAN, as we increase the number of nodes and keep the number of boosters fixed at 15\%. In the cases of 2 and 20 nodes, the effective bitrates at the gateway are 0.95kbps and 9.11kbps in \name{}, compared to 0.47kbps and 3.91kbps in LoRaWAN, respectively. Overall, such low bitrates at the LoRa gateways are due to the {\em ADR} feature of the LoRa nodes, which lets the nodes operate at higher SFs and on narrower channels. However, the bitrate at the gateway in \name{} is significantly higher than that in LoRaWAN. In the case of 20 nodes, \name{} has almost 3x higher bitrate than that in LoRaWAN.

\subsubsection{E2E Latency and Energy Consumption}\label{sec:network-latency-energy}
Figure~\ref{fig:latency_ed} shows the average E2E latency per packet at the nodes in both \name{} and LoRaWAN. In general, the packets in \name{} observe much lower E2E latency than that in LoRaWAN for all the cases as we increase the number of nodes from 2 to 20 (boosters fixed at 15\%). For 20 nodes, the average E2E latency per packet is approximately 4.5 seconds in \name{}, compared to 10 seconds in LoRaWAN. The ACK relays by the boosters play a vital role in the better performance of \name{}.
Figure~\ref{fig:energy_ed} shows the average energy consumption per packet at the nodes for both \name{} and LoRaWAN. The trend in the performances of \name{} and LoRaWAN in terms of average energy consumption per packet at the nodes is also similar to the performance trend in their average E2E latency per packet. For 20 nodes, the average energy consumption per packet is 250mJ, compared to 620mJ in LoRaWAN, thereby reducing the energy consumption per packet approximately 2.5x. Overall, all these experiments suggest that \name{} is much more suited than LoRaWAN in indoors.


\section{Related Work}\label{sec:related}
\noindent{\bf LPWAN Technologies.} Recently, many LPWAN technologies have been developed targeting licensed (e.g., cellular band), unlicensed (e.g., ISM band), and TV (e.g., white spaces) bands~\cite{lpwan_survey1, ws_survey}. LPWANs operating in the licensed band include LTE Cat M1, NB-IoT, and 5G. They require costly infrastructure and high service fees. LPWANs operating in the unlicensed band include LoRa, SigFox, RPMA, IQRF, DASH7, Telensa, Weightless-N/P, IEEE 802.11ah, IEEE 802.15.4k, and IEEE 802.15.4g. Similarly, SNOW has been developed to operate in the TV white spaces~\cite{rahmanenabling, rahman2021lpwan, rahman2020integrating, rahman2019implementation, saifullah2018low, saifullah2017enabling, saifullah2016snow}. Among these LPWANs, there is an increasing interest in LoRa from both the academic and industrial communities because of its wide adoption in an increasing number of IoT applications (e.g., smart city, smart farming, environmental monitoring, etc.)~\cite{overview1, overview2, overview3, fahmida2022real, li2021nelora, hou2022don}. Similarly, in this paper, we mainly focus on the LoRa technology.

\noindent{\bf Outdoor vs. Indoor LoRa.} 
Most existing work on LoRa focuses on evaluating and improving its performance in outdoor environment and scenarios through modeling, experiments, and simulations~\cite{germany, oliveira2017long, hosseinzadeh2017empirical, shi2019lorabee, shi2019enabling, LoRaPerformance1, fargas2017gps, khalil2018lora, LoRaSecurity3, naoui2016enhancing, LoRaSecurity2, fahmida2020long, liando2019known, xia2023xcopy}. Some measurement study in \cite{hadwen2017energy, mdhaffar2017iot, haxhibeqiri2017lora, henriksson2016indoor, manzoni2019indoor, gregora2016indoor, xu2019measurement} has recently showed the performance of LoRa in indoor environments in terms of RSSI and SNR at the gateway. To the best of our knowledge, no work has yet studied the reliability and energy-efficiency aspects of indoor LoRa at both the gateway and the LoRa nodes/sensors. In this paper, we show that only RSSI-based and SNR-based measurements (which are also the basis of outdoor measurements) may not unfold all the shortcomings that are responsible for the poor performance of LoRa gateway or nodes in indoor scenarios/applications. Additionally, through our indoor evaluation of LoRa, we propose \name{} to boost its performance indoors in terms of reliability and energy-efficiency. Note that the adoption of boosters makes our work specific to indoor scenarios since boosting a signal (by creating constructive interference) in outdoor scenarios far away may not be feasible because of the 
time difference in signal propagation from the boosters and other LoRa nodes to the gateway, and vice versa. Additionally, outdoor pathloss and shadowing effects may reduce the chances of constructive interference in \name{}.


\noindent{\bf Collision Recovery Techniques vs. \name{}.} 
Recently, several works have been proposed to recover the packets at the gateway from a collision of multiple LoRa packets~\cite{Choir, xia2019ftrack, tong2020combating, wang2019mlora, tong2020colora}. 
Choir~\cite{Choir} utilizes the distinctive hardware imperfections of the LoRa nodes. FTrack~\cite{xia2019ftrack} utilizes the distinctive signal edges of the LoRa nodes. mLoRa~\cite{wang2019mlora} applies successive interference cancellation. CoLoRa~\cite{tong2020colora} transforms the time offsets between the collided packets to frequency domain information. NScale~\cite{tong2020combating} scales up the FFT peaks of collided signals using a non-stationary signal to disentangle the collision. To the best of our knowledge, these techniques are applicable in both outdoor and indoor scenarios as long as the LoRa signal propagation is intact. However, they do not consider reliability at the LoRa nodes,
which may be critical in many scenarios (e.g., confirmed messaging for control applications). In fact, LoRaWAN may require a new operational class along with class-A/B/C to facilitate this for all the collided packets at the same frequency. 
Moreover, these techniques may not be adoptable in the commercial LoRaWAN gateways since they tend to modify/rework the LoRa physical (PHY) layer and/or signal decoding.
In \name{}, we focus on boosting the signals that suffer from severe multi-path and shadowing effects and path-loss while considering the confirmed messaging scheme of LoRaWAN. In a nutshell, \name{} is not built to handle collisions (e.g., boosters cannot decode collisions), but those collision recovery techniques are still applicable on top of \name{}. In other words, \name{} and the collision recovery techniques may complement each other.

\noindent{\bf Relaying in LoRa vs. \name{}.}
A few works have focused on relay-based performance (e.g., network coverage or lifetime) improvement in LoRa~\cite{borkotoky2021coded, tran2020two, fahmida2020long, tanjung2020oodc}. The work in~\cite{borkotoky2021coded} assumes that the gateway provides no feedback for lost packets and relay nodes do not synchronize with other nodes. Such assumptions may create duplicate packets or interference at the gateway. The works in~\cite{tran2020two, fahmida2020long, tanjung2020oodc} improve the network coverage and/or lifetime by enabling multi-hop communication, in which the relay nodes (e.g., those closer to the gateway or have more battery budgets in energy-harvesting networks) forward packets to the gateway. They, however, do not ensure
that the closest hop to the gateway delivers packets reliably. Adopting these techniques may not thus improve performance in indoor LoRa. In contrast, \name{} ensures reliability by suppressing duplicates and interference at the gateway.

\noindent{\bf Backscattering in LoRa vs. \name{}.}
\revise{The works on LoRa backscattered systems develop algorithms that require customized hardware to facilitate the encoding, detection, and decoding of backscattered LoRa signals~\cite{peng2018plora, guo2020aloba, guo2021efficient, hessar2019netscatter, jiang2021sense, jiang2021long, talla2017lora, guo2022saiyan, varshney2017lorea}. Consequently, these techniques may not be readily adoptable in the commercial LoRaWAN gateways or end-devices. On the other hand, the goals and design of \name{} are fundamentally different and require no hardware modification. The body of work on LoRa backscattered systems, however, may adopt  \name{} to make them more reliable and energy-efficient in indoor environments since backscattered passive LoRa signals are much weaker compared to the active ones.}


\noindent{\bf Non-destructive vs. Constructive Interference in LoRa.} 
A few works have experimentally demonstrated that LoRa can decode two concurrent packets with 1dB difference in signal strength if their start-of-transmissions do not differ more than three LoRa symbol periods~\cite{overview1, liao2017multi, ma2020poster}. Three LoRa symbol duration ranges between 768$\mu$s and 98.3ms depending on different spreading factors and channel bandwidths of LoRa. Such a concurrency is termed as {\em non-destructive} transmissions or interference in the literature, which is also a basis of multi-hop communication in LoRa~\cite{overview1, mai2020multi, sartori2017enabling}. In \name{}, we reinforce the strengths of certain packets via {\em constructive interference} and then decode them. This enables the decoding of certain packets (e.g., packets of the nodes that suffer most) from {\em many} non-decodable concurrent ones with different signal strengths and temporal displacements.

\section{Conclusions}\label{sec:conclusion}
In this paper, we have proposed a link-layer protocol called \name{} to boost the reliability at the gateway and energy-efficiency at the end-devices of indoor LoRa networks. To do this, we first extensively evaluated the performance of the LoRa MAC protocol -- LoRaWAN -- indoor and showed that the reliability at the gateway was as low as 62\%. Also, the number of retransmissions per packet was as high as 4 to 5. \revise{In \name{}, we boosted the reliability and energy-efficiency in the network by creating constructive interference of the packets at the gateway with specific timing requirements (analyzed both empirically and mathematically)} and relaying missing acknowledgments to the end-devices. Our extensive experiments using \name{} showed that the reliability at the gateway in indoor LoRa networks increased from 62\% to 95\% while the end-devices operated 2.5x efficiently in terms of energy, thereby demonstrating the feasibility of \name{}.


\bibliographystyle{IEEEtran}
\balance
\bibliography{IEEEabrv,lorafull}

\vfill

\end{document}